# Enriching and Characterizing T-Cell Repertoires from 3' Barcoded Single-Cell Whole Transcriptome Amplification Products

Authors: Tasneem Jivanjee[1,2,3,4,#], Samira Ibrahim[1,2,3,4,#], Sarah K. Nyquist[1,2,3,4,5,6], G. James Gatter[1,2,3,4], Joshua D. Bromley[1,2,3,4,7], Swati Jaiswal[8], Bonnie Berger[6,9], Samuel M. Behar[8], J. Christopher Love[2,3,4,10,*], Alex K. Shalek[1,2,3,4,5*]

[#]These first authors contributed equally to this work.
*These senior authors contributed equally to this work.

Affiliations:

1. Institute for Medical Engineering & Science and Department of Chemistry, Massachusetts Institute of Technology, Cambridge, MA

2. Koch Institute for Integrative Cancer Research, Massachusetts Institute of Technology, Cambridge, MA

3. Ragon Institute of MGH, MIT, and Harvard, Cambridge, MA

4. Broad Institute of MIT and Harvard, Cambridge, MA, Massachusetts Institute of Technology, Cambridge, MA

5. Program in Computational and Systems Biology, Massachusetts Institute of Technology, Cambridge, MA, USA

6. Computer Science and Artificial Intelligence Laboratory, Massachusetts Institute of Technology, Cambridge, MA, USA.

7. Microbiology Graduate Program, Massachusetts Institute of Technology, Cambridge, MA 02139, USA

8. Department of Microbiology and Physiological Systems, University of Massachusetts Medical School, Worcester, Massachusetts, USA

9. Department of Mathematics, Massachusetts Institute of Technology, Cambridge, MA

10. Department of Chemical Engineering, Massachusetts Institute of Technology, Cambridge, MA

Corresponding authors: **J. Christopher Love: [clove@mit.edu](mailto:clove@mit.edu); Alex K. Shalek: [shalek@mit.edu](mailto:shalek@mit.edu)**

**Abstract**

Antigen-specific T cells play an essential role in immunoregulation and many diseases such as cancer. Characterizing the T cell receptor (TCR) sequences that encode T cell specificity is critical for elucidating the antigenic determinants of immunological diseases and designing therapeutic remedies. However, methods of obtaining single-cell TCR sequencing data are labor and cost intensive, typically requiring both cell sorting and full length single-cell RNA-sequencing (scRNA-seq). New high-throughput 3' cell-barcoding scRNA-seq methods can simplify and scale this process; however, they do not routinely capture TCR sequences during library preparation and sequencing. While 5' cell-barcoding scRNA-seq methods can be used to examine TCR repertoire at single-cell resolution, doing so requires specialized reagents which cannot be applied to samples previously processed using 3' cell-barcoding methods.

Here, we outline a method for sequencing *TCRα* and *TCRβ* transcripts from samples already processed using 3' cell-barcoding scRNA-seq platforms, ensuring TCR recovery at a single-cell resolution. In short, a fraction of the 3' barcoded whole transcriptome amplification (WTA) product typically used to generate a massively parallel 3' scRNA-seq library is enriched for TCR transcripts using biotinylated probes, and further amplified using the same universal primer sequence from WTA. Primer extension using TCR V-region primers and targeted PCR amplification using a second universal primer results in a 3' barcoded single-cell *CDR3*-enriched library that can be sequenced with custom sequencing primers. Coupled with 3' scRNA-seq of the same WTA, this method enables simultaneous analysis of single-cell transcriptomes and TCR sequences which can help interpret inherent heterogeneity among antigen-specific T cells and salient disease biology. The method presented here can also be adapted readily to enrich and sequence other transcripts of interest from both 3' and 5' barcoded scRNA-seq WTA libraries.

**Key words:** Single-cell RNA-sequencing, scRNA-seq, Seq-Well S$^3$, T-cell receptor repertoire profiling, Gene expression, Targeted Enrichment, single-cell TCR sequencing

## 1.    Introduction

Antigen-specific T cells play a critical role in immunoregulation and various diseases such as cancer and autoimmune disorders *(1–4)*. Each T cell possesses a genetically encoded T cell receptor (TCR); at any given moment, the repertoire of all TCRs and the T cell states to which each is tied define which specific antigens a host can identify and how it will respond to them, respectively, informing key aspects of adaptive immunity *(5)*. Critically, a host's overall TCR repertoire is not fixed, but rather can change in response to internal and external factors, such as spontaneous mutations or infection, respectively *(6)*.  Thus, sequencing this region is essential for characterizing TCR diversity. Yet, the diversity of these TCRs can also present a challenge when it comes to precisely analyzing the repertoire and its relation to cellular phenotypes *(7)*. To improve the accuracy of identifying key T cell subpopulations, methods that can link genetically encoded receptor identity and phenotype to adaptive immune responses are needed *(8)*.

The TCR itself is a heterodimer of two chains (αβ or γδ). The uniqueness of each T cell's TCR is the result of recombination events occurring between the genes encoding for the TCR α and β (or γ and δ) variable (V), diversity (D), and joining (J) regions. For the TCR α chain, recombination results in one of myriad VJ sequences; for the β, it yields a VDJ. The unique part of each α and β

TCR chain is contained within its complementarity-determining region 3 (CDR3). This region includes the V(D)J recombination junction as well as the part of the receptor that binds to major histocompatibility complex (MHC) molecules on antigen-presenting cells *(5)*.

To investigate the sequences of the TCR α and β chains belonging to specific T cells, a few methods have emerged. One approach is bulk genome sequencing of the entire TCR repertoire. Although this strategy (and related ones) enables comprehensive population-level TCR analysis, it mainly provides insight into overall TCRα and TCRβ usage rather than revealing distinct single-cell TCRα-TCRβ pairs (though this can be computationally inferred); similarly, it does not allow exploration of gene expression at the single-cell level, hindering understanding of the relationship between TCR clonotype (a cell's specific α, β pair) and biological function *(6)*. Another analysis method uses tetramers to tag antigen-specific T cells. Here, fluorescently- (or oligonucleotide-) labeled MHC-peptide tetramers are used to mark antigen-specific T cells for subsequent analysis *(9)*. This approach can be combined with cell sorting methods to identify the clonotypes that bind to the specific tetramer. Specifically, tetramer-labeled cells are single-cell sorted and then each cell's αβ TCR pair is amplified by targeted PCR and determined by sequencing *(9, 10)*. To link clonotype to phenotype, this initial targeted PCR can be replaced by whole transcriptome amplification (WTA); after, the sequences of the *TCRα* and *TCRβ* can be reconstructed computationally from full length single-cell RNA-sequencing (scRNA-seq) data or probed directly by targeted library generation and sequencing *(11)*. However, these plate-based methods have limited throughput due to labor, time and cost constraints *(6, 12, 13)*.

To scale phenotypic profiling, newer high-throughput scRNA-seq methods have been developed that rely on early cell barcoding. Early tagging of each cell's mRNAs with unique nucleic acid sequences allows many cells to be processed at once while still enabling computational grouping of the transcripts belonging to each cell *(14)*. While these methods may be useful for high-throughput single-cell gene expression profiling, most 3' barcoded scRNA-seq methods do not provide information on the variable region of the α and β TCR genes—particularly the complementarity-determining region 3 (*CDR3*) regions which lie closer to the 5' end of the TCR transcripts and are effectively lost in most 3' library preparation methods which generate short fragments (~300-500 bp) for sequencing (Fig. 1). In other words, 3' scRNA-seq methods (such as Drop-seq, Seq-Well S$^3$, 10x 3', and inDrops, or similar spatial methods such as Slide-Seq) typically do not return fragments for sequencing with both CDR3 sequences and the 3' cell barcode and unique molecular identifier (UMI). This information is necessary to determine the identify of a cell's α and β TCR sequences and to link them to a specific cell identifier, respectively, so that they may be tied to a specific phenotypic state (as determined by the other transcripts that the cell expresses) *(15–17)*. While targeted amplification using a primer against the TCR constant region works for 5' cell barcoded WTA products, such as those created using the 10x 5' v2 kit, it requires special reagents (a TCR sequencing kit). Furthermore, it cannot be applied to samples previously processed using 3' cell barcoding methods since doing so will eliminate the 3' cell barcodes and UMIs, ultimately obscuring single-cell resolution *(7)*. An alternative approach is to couple short reads for gene expression on one sequencing platform (e.g., Illumina) with long reads from another (e.g., Nanopore) to measure phenotype and genotype respectively, as in RAGE-seq, but this approach requires two sequencing platforms and is difficult to scale *(7)*.

Here, we report a method for simultaneously sequencing the transcriptome and *TCRα* and *TCRβ* sequences of T cells using Seq-Well S[3], a portable, low-cost platform for massively parallel 3' scRNA-seq, in combination with a TCR target enrichment strategy (*18*). Seq-Well S[3] enables confinement of single cells—whether from human, mouse, non-human primates (NHPs), or the like—with barcoded poly(dT) mRNA capture beads in a PDMS array of sub-nanoliter wells sealed with a semi-permeable membrane; this allows efficient cell lysis through fast solution exchange and improves transcript capture (*18, 19*). In Seq-Well S[3], we obtain a 3' barcoded WTA product as outlined in the online Seq-Well S[3] protocol (http://shaleklab.com/wp-content/uploads/2019/07/SeqWell-S3-Protocol.pdf). The method as described below also works for enriching TCR repertoires from Drop-seq and Slide-Seq libraries, and can be adapted with minor modification to other 3' barcoded WTA products, such as those generated using 10x 3' kits, or, with additional changes, to 5' barcoded WTA products (*15, 16, 18*).

From the amplified WTA product, a fraction of the material is used to generate a 3' barcoded whole transcriptome scRNA-seq library. Subsequent sequencing enables mRNA expression quantification from thousands of cells at once. Another fraction of the WTA product is used to capture and enrich *TCRα* and *TCRβ* transcripts contained within the WTA product, and then targeted sequencing is performed on those specific transcripts. Collectively, this approach recovers both the full transcriptome and the TCR sequence of each T cell, enabling direct linkage of T cell clonotype to T cell phenotype. While this chapter will focus on TCR transcript enrichment and sequencing, other sequences of interest, such as oncogenes or viral nucleic acids, can be similarly enriched by appropriately adapting the various primers provided (*20, 21*). We also note that this method can be applied to multiple species, as evidenced in the notes regarding primer design (*see* **Notes 1, 2**).

More specifically, the single-cell TCR sequencing protocol presented here utilizes biotinylated probes targeting the *TCRα* and *TCRβ* constant regions (Fig. 2). After affinity capture, the remaining WTA product, which is enriched for *TCRα* and *TCRβ* complementary DNAs (cDNAs), is subsequently amplified using the same universal primer site (UPS) employed to obtain the original WTA library. Through primer extension, a second 5' universal primer site (UPS2) linked to the TCRα and TCRβ variable regions is then used to modify the pool of TCR nucleic acids. Finally, PCR amplification using a pair of primers that bind to the UPS2 site as well as an extended version of the UPS yields an enriched 3' barcoded single-cell TCR sequencing library. This CDR3-enriched library can be sequenced on an Illumina MiSeq (or NextSeq) using custom sequencing primers specific to the constant regions of the *TCR* genes to recover the variable region of the *TCRα* and *TCRβ* chains.  Afterward, these single-cell *TCRα* and *TCRβ* sequences can be matched to whole transcriptome profiles obtained via traditional 3' scRNA-seq methods based on shared 3' cell barcodes *(7)*. In short, sequencing reads are aligned to known TCR sequences and the V region, J region, and CDR3 sequences are identified for each UMI using a computational pipeline implemented in Python. This alignment pipeline returns a table containing these annotations and their corresponding bead barcodes. TCR variable regions may then be matched to corresponding cells from the Seq-Well S[3] output by matching the TCR output barcodes to those in the corresponding Seq-Well S[3] library.

Below, we provide a detailed walk through of both the experimental and computational pipeline for capturing and annotating single-cell *TCRα* and *TCRβ* sequences.

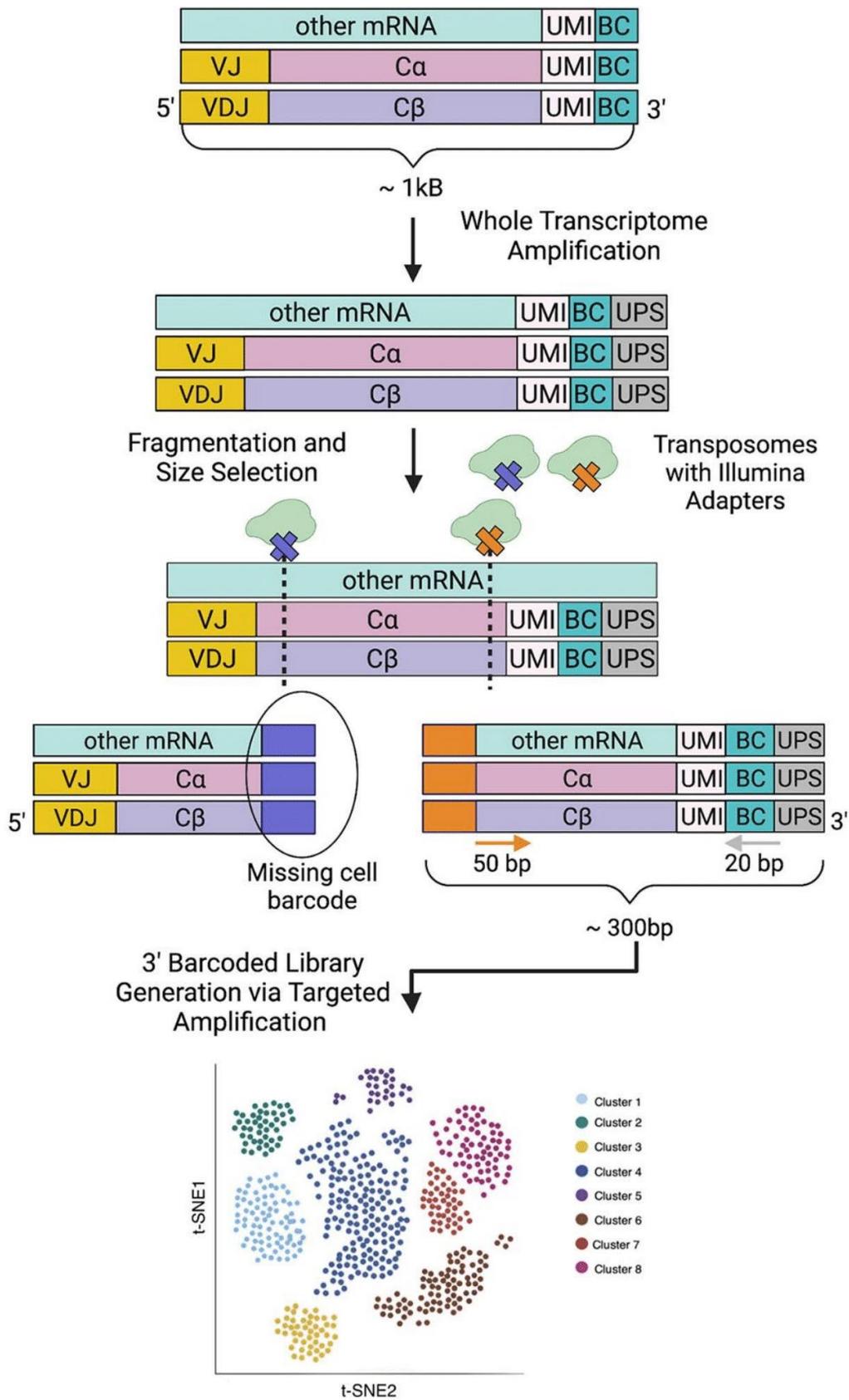

**Fig. 1** Using Seq-Well S³, a 3' scRNA-seq method, captured mRNA transcripts are amplified via whole transcriptome amplification (WTA). WTA products, including TCR transcripts, undergo fragmentation, tagmentation, and size selection, resulting in a 3' barcoded library of roughly 300-500 bp size fragments that can be sequenced. For captured TCR transcripts, this process significantly minimizes the chances of obtaining the CDR3 reads in the final library, as this region is approximately 1 kB away from the 3' barcoded region.

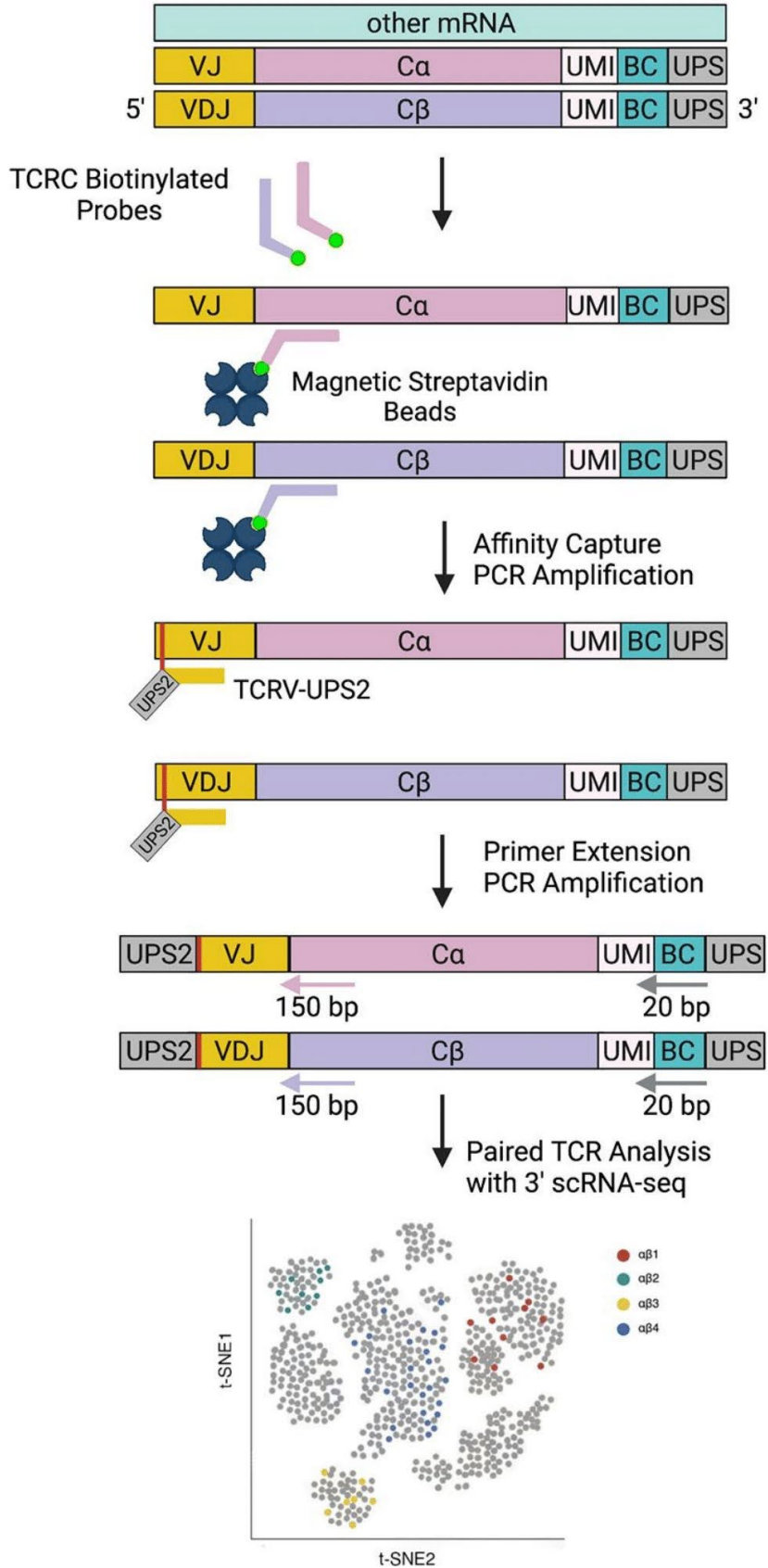

**Fig. 2** The amplified WTA product which contains the *TCRα* and *TCRβ* transcripts undergoes a 3' cell barcoded library preparation as outlined in the Seq-Well S$^3$ protocol. Affinity capture with species-specific biotinylated oligonucleotides (biotin shown in green) targeting TCR constant regions enriches 3' barcoded TCR transcripts. A sequencing library of these products is then prepared via primer extension with species-specific V-region primers containing UPS2 handles and subsequent PCR amplification with UPS2 and an extended version of UPS. Finally, species-specific custom sequencing primers are used for sequencing the CDR3 region on an Illumina MiSeq and analyzed together with the full transcriptomic data.

## 2. Materials

### 2.1. Seq-Well S$^3$ TCR Recovery Materials

1. xGen Hybridization and Wash Kit (IDT, USA), store at 4 ºC and -20 ºC respectively.

2. Biotinylated oligonucleotide probes for TCR enrichment, store at -20 ºC (*see* **Note 2**).

3. UPS, stock concentration is 100 μM, store at -20 ºC.

<p align="center">5'AAGCAGTGGTATCAACGCAGAGT</p>

4. Nuclease-free water.

5. DNA Lo-Bind tubes, 1.5 mL.

6. PCR strip tubes.

7. 15 mL conical tube.

8. Pipettes (P2, P20, P200, P1000).

9. Low retention, sterile, filtered pipet tips (20 μL, 200 μL, 1000 μL).

10. Kapa Hifi Hotstart Readymix (2✕), store at -20ºC.

11. AMPure XP beads (Beckman Coulter, USA), store at 4 ºC.

12. Reagent reservoirs.

13. Ethanol, pure (200 proof, anhydrous).

14. High Sensitivity D5000 Reagents (Agilent, USA), store at 4 ºC.

15. High Sensitivity D5000 ScreenTape (Agilent, USA), store at 4 ºC.

16. TCRV-UPS2 TCRα (*see* **Note 1** for primer design).

17. TCRV-UPS2 TCRβ.

18. 96-well plate.

19. UPS-mod-N50x, stock concentration is 1 mM, store at -20 ºC (*see* **Note 3** for option for using Illumina barcodes).

5'AATGATACGGCGACCACCGAGATCTACACGCCTGTCCGCGGA AGCAGTGGTATCAACGCAGAGT*A*C

20. UPS-N70x, stock concentration is 1 mM, store at -20 ºC.

5'CAAGCAGAAGACGGCATACGAGATGTCTCGTGGGCTCGG

21. Custom sequencing primer, stock concentration is 100 μM, store at -20 ºC (*see* **Note 4**).

22.  Seq-Well S$^3$ Sequencing Primer, stock concentration is 100 μM, store at -20 ºC.

           5'GCCTGTCCGCGGAAGCAGTGGTATCAACGCAGAGTAC

23. Illumina MiSeq 150 cycle kit.

## 2.2.　Equipment

1.  Thermocycler.

2.  Vortex.

3.  Magnet for PCR strip tubes.

4.  DynaMag-2 Magnet.

5.  Thermomixer.

6.  Mini centrifuge.

7.  Centrifuge.

8.  Plate/PCR strip vortex.

9.  Qubit Fluorometer.

10. Kapa Library Quantification qPCR.

11. Advanced Analytical Fragment Analyzer.

12. Illumina MiSeq.

## 2.3.　Computer

A computer capable of running Docker and access to Google Cloud.

## 2.4.　Data Matrix Files

Digital expression matrices from Seq-Well S³ libraries including cell barcodes for each library.

## 3. Methods

Carry out all temperature sensitive procedures at suitable temperature as illustrated (*see* **Note 5**).

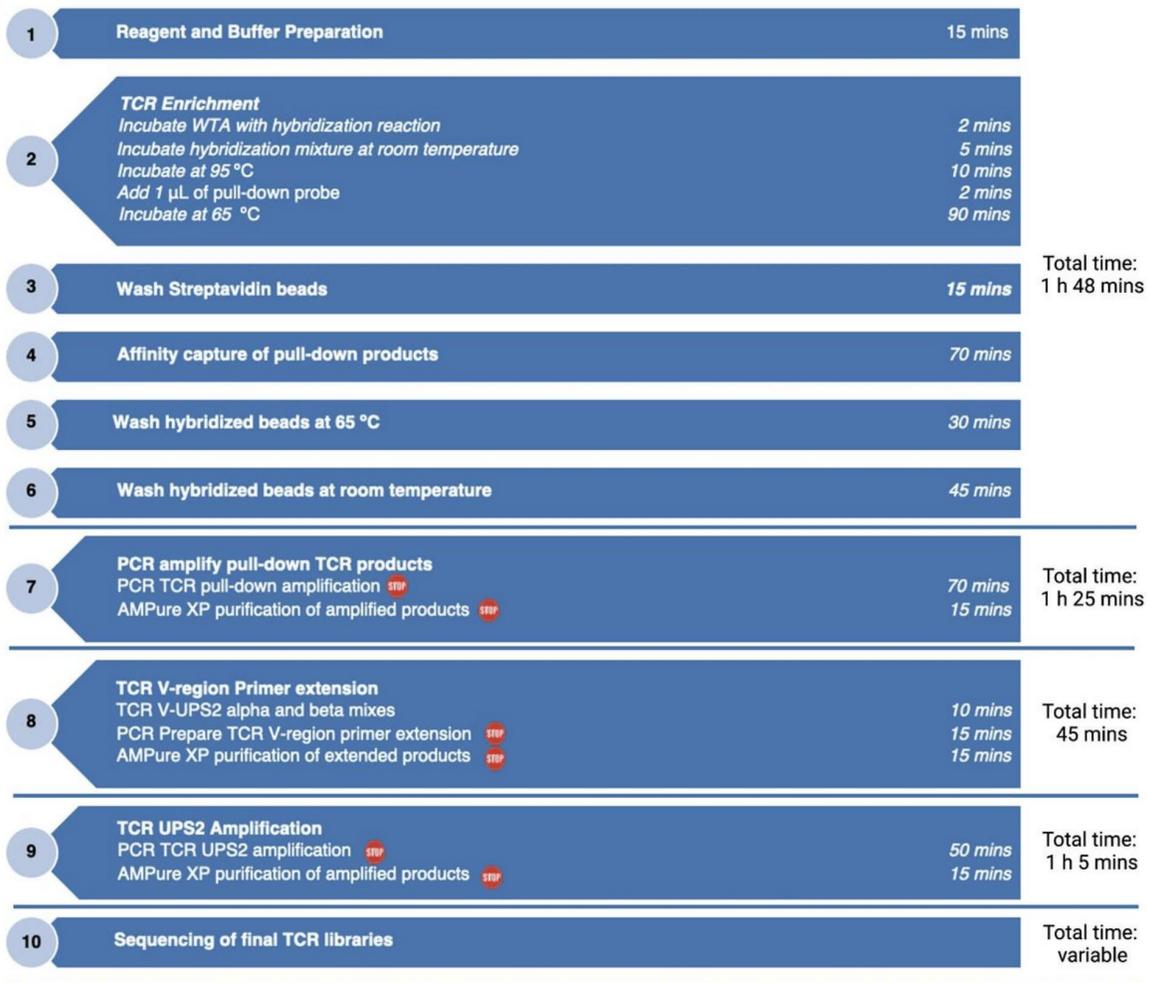

**Fig. 3** Workflow depicting the full TCR enrichment protocol with total time of each step specified as well as stopping points in the protocol.

## 3.1.    TCR Enrichment from 3' Barcoded Single-cell Libraries

### 3.1.1.    Reagent Preparation

1.  Dilute Wash Buffer 1, Wash Buffer 2, Wash Buffer 3, and Stringent Wash Buffer to working concentrations. Aliquot in 0.2 mL PCR strip tubes and store at 4 ºC.

2.  Dilute Bead Washing Buffer to working concentration. Aliquot in a 50 mL conical tube and store at 4 ºC.

3.  Allow Wash Buffer 1, Wash Buffer 2, Wash Buffer 3, Stringent Wash Buffer, and Bead Washing Buffer to come to room temperature before beginning (*see* **Note 5**).

### 3.1.2.    TCR Enrichment

1.  TCR enrichment is performed with whole transcriptome amplified (WTA) products which are generated by performing Seq-Well S$^3$/Drop-seq/Slide-seq methods.

2.  Set up a 12.5 µL hybridization reaction with 3.5 µL of WTA product:

| Master Mix 1 (per sample/reaction) | 1x (µL) |
|---|---|
| UPS (50 µM) | 0.8 |
| Cot-1 DNA | 0.5 |
| 2× Hybridization Buffer | 8.5 |
| xGen Buffer Enhancer | 2.7 |
| Total volume | 12.5 |

**Table 1** Reagents necessary for master mix 1 of TCR enrichment.

3. Mix 3.5 µL of WTA product with the hybridization reaction mix in 0.2 mL PCR strip tubes.

4. Incubate the tube strip at room temperature for 5 mins. Then, incubate at 95 ºC for 10 mins (*see* **Note 6**).

5. Remove mixture to room temperature and add 1 µL of pull-down probe mix (*see* **Note 2**). Vortex and centrifuge the strip tubes to mix and collect the liquid respectively.

6. Incubate the mixture at 65 ºC for 1 hr to allow for hybridization of pull-down probes to the WTA libraries.

### 3.1.3. Wash Streptavidin Beads

1. Aliquot 50 µL of streptavidin beads per sample into a 1.5 mL microcentrifuge tube. Place the tube on a DynaMag magnetic stand and allow for the beads to pellet.

2. Remove supernatant and add equal volume of bead wash buffer (BWB) into the tube. (*see* **Section 3.1.1**)

3. Vortex the tube to mix, and place back on the magnetic stand for beads to pellet.

4. Repeat steps 2-3

5. Take the tube off the magnet and add an equal volume of bead wash buffer.

6. Aliquot 50 μL (per sample) of the mixture into PCR strip tubes.

7. Place the tubes on a PCR strip magnet and remove the supernatant once the hybridized mixture from **Section 3.1.2** is ready (*see* **Note 7**).

### 3.1.4.  Affinity Capture of Pull-down Products

1. Add the hybridized mixture from **Section 3.1.2** into the prepared streptavidin beads.

2. Vortex and spin down briefly to mix and collect liquid respectively.

3. Incubate mixture at 65 ºC for 45 min. Intermittently vortex the mixture every 10 min to ensure the beads are suspended in the mixture (*see* **Note 8**).

4. During the incubation, preheat aliquoted Wash Buffer 1 (WB1) and Stringent Wash Buffer (SWB) at 65 ºC (*see* **Note 9**). Preheat at

least 100 µL/sample of WB1 and 400 µL/sample of SWB (2x200 µL).

### 3.1.5. Wash hybridized beads at 65 degrees

1. While the strip tubes are at 65 ºC, add 100 µL of heated WB1 into each tube, pipette up and down gently, scraping the beads from the side of the tubes.

2. Place the PCR strip tube on the magnet, aspirate supernatant, and take tubes off the magnetic stand and return to 65 ºC (*see* **Note 10**).

3. While the strip tubes are at 65 ºC, add 200 µL of heated SWB into each tube, and pipette up and down gently resuspend the bead pellets.

4. Incubate mixture at 65 ºC for 5 min.

5. Repeat steps 2-4.

6. Aspirate final wash of SWB.

### 3.1.6. Wash hybridized beads at Room Temperature

1. Add 200 µL of room temperature WB1 into each reaction, pipette up and down to resuspend beads.

2. Vortex for 2 min to mix, spin briefly to collect liquid, and place on a magnetic stand.

3. Aspirate and add 200 µL of room temperature WB2, pipette up and down to resuspend beads.

4. Vortex for 1 min to mix, spin to collect liquid, and place on a magnetic stand.

5. Aspirate and add 200 µL of room temperature WB3, pipette up and down to resuspend beads.

6. Vortex for 30 seconds to mix, spin to collect liquid, and place on a magnetic stand.

7. Aspirate and resuspend the beads in 20 µL of nuclease-free water.

### 3.1.7. PCR Amplify Pull-down TCR Products

1. Set up a 25 µL of PCR reaction with hybridized mixture. For each mixture, perform 5 PCR reactions, using 2 µL of mixture in each reaction (using a total of 10 out of 20 µL):

| TCR pull-down amplification PCR system | 1x (µL) |
|---|---|
| Kapa Hifi Hotstart Readymix (2✕) | 12.5 |
| UPS (10 µM) | 2 |
| Nuclease-free water | 8.5 |
| Hybridized mixture | 2.5 |
| Total volume | 25 |

**Table 2** Reagents necessary for TCR pull-down amplification PCR system.

2. Run the PCR using the following conditions:

| Cycles | Temperature ( ℃ ) | Time |
|--------|-------------------|------|
| 1 | 95 | 3 minutes |
| 25 | 98 | 40 seconds |
| | 67 | 20 seconds |
| | 72 | 1 minute |
| 1 | 72 | 5 minutes |

**Table 3** Outline of PCR program for TCR pull-down amplification.

3. After amplification, pool all 5 PCR reactions (20 μL for each sample) into a single tube or well in a 96 well plate (for 125 μL total).

### 3.1.8.  Purify TCR pull-down products and assess quality

1. Add 32 μL AMPure XP reagent to pooled PCR product from previous step (0.64✕ volumetric ratio); mix by pipetting (*see* **Note 11**).

2. Allow the tubes to incubate for 5 min at room temperature.

3. Place the tube on the rack for 3 min and remove the supernatant.

4. Add 100 μL of fresh 80% ethanol to the tube. After addition, shift the tube on the magnet 6 times to allow beads to pass through the ethanol solution to the opposite side of the tube.

5. Remove 80% ethanol wash solution.

6. Repeat steps 3-5 twice.

7. After the third ethanol wash, remove any residual ethanol and allow beads to air dry for 4 min (*see* **Note 12**).

8. Resuspend beads in 15 μL nuclease-free water.

9. Place the tubes on the magnet stand and transfer the 15 μL supernatant, which contains the eluted full-length TCR alpha and beta products, to a fresh 96-well plate.

10. Assess quality of the enrichment via Advanced Analytical Fragment Analyzer or running an Agilent D5000 High Sensitivity Screentape (*see* **Note 13**) (Fig. 4).

11. Determine concentration using Qubit Fluorometer or Tape report.

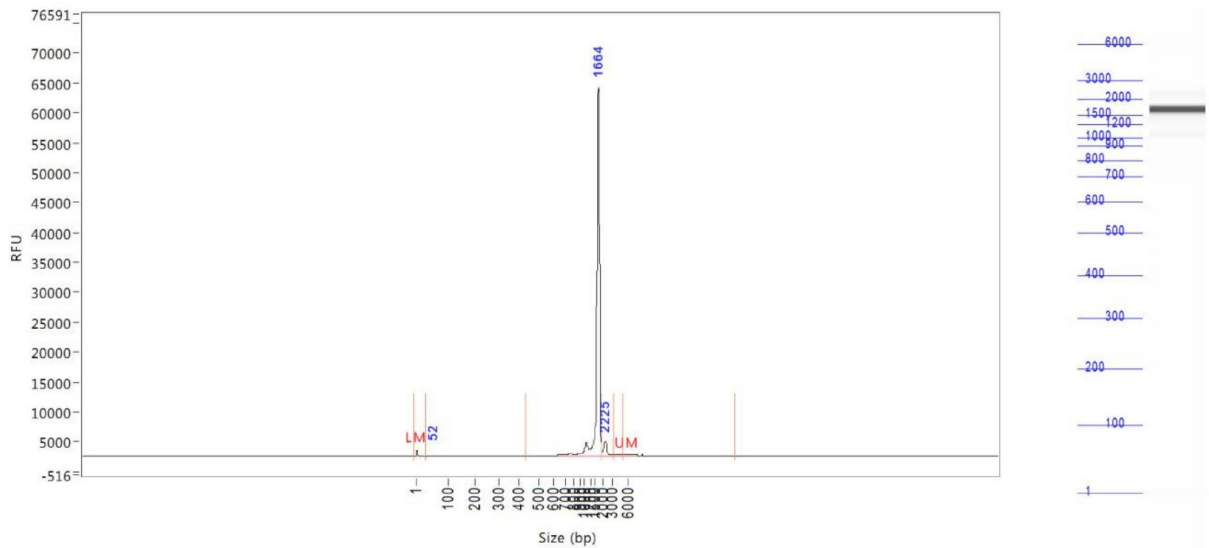

**Fig. 4** Expected size distribution of amplified TCRα and TCRβ pull-down products (*see* **Note 14**).

### 3.2. TCR V-region Primer Extension

1. Prepare separate TCRV-UPS2 alpha and beta mixes by equimolar pool each of the primers and dilute as necessary to 10 µM (total concentration) (*see* **Note 15**).

2. Set up a 25 µL of PCR reaction with enriched full-length TCR products. Make two separate reactions (one for TCRα and one for TCRβ chain).

| TCR V-region primer extension PCR system | 1x (µL) |
|---|---|
| Kapa Hifi Hotstart Readymix (2✕) | 12.5 |
| TCRV-UPS2 TCRα and TCRβ (10 µM) | 2.5 |
| Nuclease-free water | 6 |
| Hybridized mixture | 4 |
| Total volume | 25 |

**Table 4** Reagents necessary for TCR V-region primer extension PCR system.

3. Perform the primer extension using the following conditions:

| Cycles | Temperature ( ℃ ) | Time |
|---|---|---|
| 1 | 95 | 5 minutes |
| 1 | 55 | 30 seconds |
| 1 | 72 | 2 minutes |

**Table 5** Outline of PCR program for TCR V-region primer extension.

4. After extension is complete, add 25 μL of water into each reaction, to bring the total to 50 μL.

5. Purify using AMPure XP beads as previously described in 3.1.8. Elute in 15 μL of water.

**3.3.    TCR UPS2 amplification:**

1. Set up a 25 μL of PCR reaction with 2.5 μL of TCR primer extension product from 3.2. Split each sample into 4 PCR reactions (*see* **Note 16**).

| TCR UPS2 amplification PCR system | 1x (μL) |
|---|---|
| Kapa Hifi Hotstart Readymix (2✕) | 12.5 |
| UPS-mod-N50x (10 μM) | 0.5 |
| UPS2-N70x (10 μM) | 0.5 |
| Nuclease-free water | 9 |
| TCR primer extension product | 2.5 |
| Total volume | 25 |

**Table 6** Reagents necessary for TCR UPS2 amplification PCR system.

2. Run the PCR using the following conditions (*see* **Note 17**):

| Cycles | Temperature ( °C ) | Time |
|---|---|---|
| 1 | 95 | 2 minutes |
| 9-18 cycles | 95<br>60<br>72 | 30 seconds<br>30 seconds<br>1.5 minutes |
| 1 | 72 | 5 minutes |

**Table 7** Outline of PCR program for TCR UPS2 amplification.

| Concentration of products from 3.1. (ng/µL) | Recommended cycle number for TCRβ | Recommended cycle number for TCRα |
|---|---|---|
| < 1 | 16-18 | 17-18 |
| 1-10 | 14 | 15 |
| 10-20 | 13 | 14 |
| 20-30 | 12 | 13 |
| 30-40 | 11 | 12 |
| >40 | 9-10 | 9-11 |

**Table 8** Recommended number of cycles for TCR UPS2 amplification based on concentration estimation of amplified pull-down products. Precise cycles will need to be tested and adjusted for each sample.

3. Take 12.5 µL of each reaction and pool for each sample for a final volume of 50 µL per sample.

4. Purify using AMPure XP beads as previously described in 3.1.8. Elute in 15 µL of water (*see* **Note 18**).

5. Assess size distribution via Advanced Analytical Fragment Analyzer or running an Agilent D5000 High Sensitivity Screentape (Fig. 5).

6. Determine final concentration using Kapa Library Quantification (*see* **Note 19**).

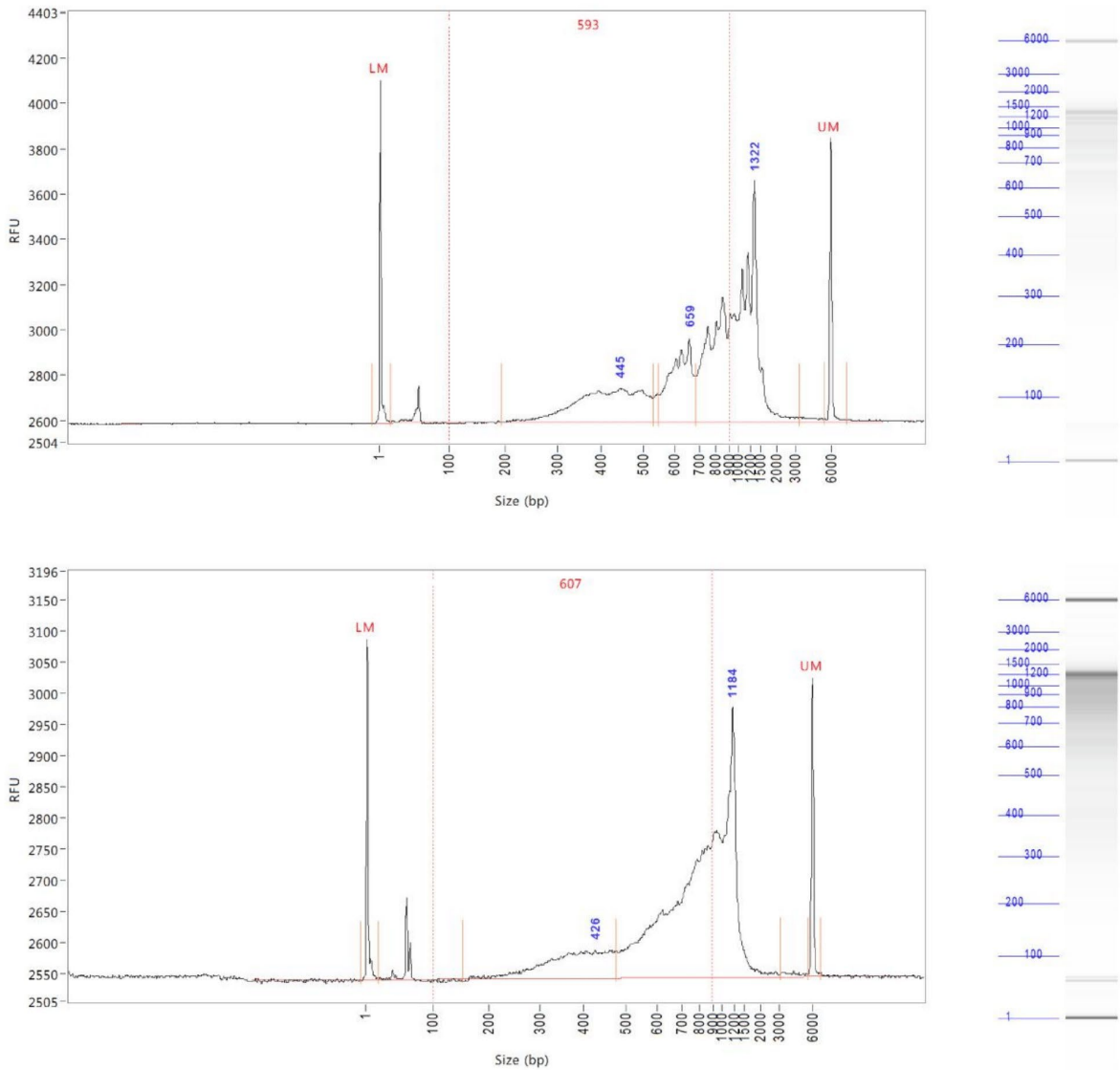

**Fig. 5** Expected size distribution of UPS2 amplified TCRα (top) and TCRβ products (bottom). TCRα products are expected to be longer than TCRβ products (*see* **Note 14**).

### 3.4. Sequencing of final TCR libraries:

Final TCRα and TCRβ libraries can be pooled based on qPCR concentrations and sequenced on Illumina Miseq using a 150-cycle kit (*see* **Note 20**). TCRα and TCRβ libraries may be pooled or sequenced separately. Sequencing primers are used at a final concentration of 2.5 μM (*see* **Note 4**). Species-specific TCR custom sequencing primers are used for Read 1 and the Seq-Well S[3] Sequencing Primer is used for index 1. We aim for $8\text{-}12 \times 10^6$ pass filter reads per lane, a cluster density of roughly $450,000/\text{mm}^2$ (*see* **Note 21**). Based on the whole-transcriptome data, we allot ~6,000 T cells per lane.

The sequencing structure is as follows (see **Note 22**):

Read 1: 150 bp

Index 1: 20 bp

### 3.5. Analysis

After sequencing, computational analysis is required to identify TCRs and connect these with their corresponding cell barcodes. If no Illumina indices are used, the final FASTQ will contain TCR sequences in Read 1 and the cell barcode with UMI in Index 1. The Illumina bcl2fastq software may be run without barcodes in the sample sheet and a single-ended FASTQ file for the sequenced pool will be output. If Illumina indices are used, see **Section 3.5.1** for demultiplexing instructions. Once FASTQ files are obtained, the TCR alignment and CDR3 identification pipeline should be run (*see* **Section 3.5.2**).

### 3.5.1. Demultiplexing with Illumina indices

When Illumina indices are included in the UPS-mod-N50x primer, a sample sheet with these indices should be used as input to bcl2fastq (https://github.com/ShalekLab/tcrgo/blob/master/examples/SampleSheet_indexed_run_example.csv). When running bcl2fastq to generate outputs for the TCRGO pipeline, it is recommended to generate paired-end FASTQ files. This can be done with the '--create-fastqs-for-index-reads' parameter and with the '--use-bases-mask' set to 'Y140,Y20,I8' in bcl2fastq. Note that the output of this run will be paired ended fastqs for each barcoded sample where Read 1 contains the 140 bp TCR read and read 2 contains the 20 bp index 2 read with the Seq-Well S$^3$ bead barcode and UMI. At this stage, both FASTQ files represent the reverse complement of the sequences (Fig. 7A), and this format is the expected input of the following pipeline.

### 3.5.2. TCR alignment and CDR3 identification pipeline

The computational Seq-Well S$^3$ TCR alignment pipeline, TCRGO, identifies and quantifies the TCR regions associated with each cell barcode in the sequencing data resulting from TCR enrichment protocol described in Tu et al. 2019 *(7)*. This pipeline is available at www.github.com/ShalekLab/tcrgo. We have reengineered the Seq-Well S$^3$ TCR recovery pipeline originally presented by Tu et al. to make significant enhancements to the user and developer experience while delivering comparable results. Chiefly, we have improved the performance, reporting, tunability, portability, and maintainability of the computational pipeline in a new Python program, TCRGO. This pipeline consists of four main steps: 1. preprocessing and alignment; 2. filtering using alignment information; 3. CDR3 recovery; and 4. summarizing the data (Fig. 6).

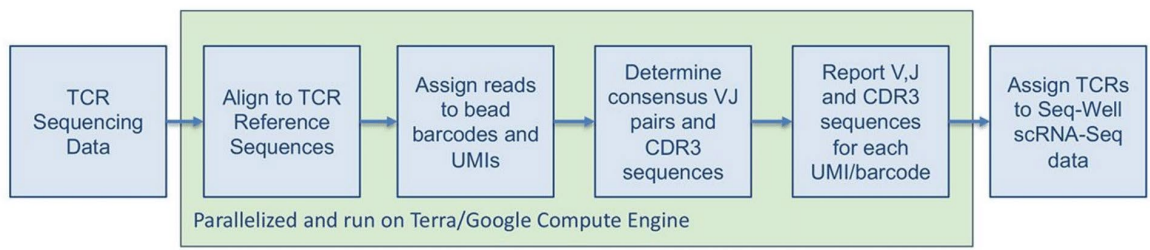

**Fig. 6** Overview of TCRGO workflow. Pipeline can be run locally or run on remote servers. Boxed section is parallelized and optimized to run on the Terra.bio platform.

**Running the pipeline**

The TCRGO pipeline is available as a Docker image, hosted on Dockstore at https://dockstore.org/workflows/github.com/ShalekLab/tcrgo/TCRGO, thus is quickly deployable in cloud computing environments. This pipeline is specifically optimized for use on the Terra.bio platform, a wrapper around Google Compute Engine for computational biology pipelines. To run this pipeline on the Terra platform, FASTQ files should be uploaded to Terra and added to a Terra data table. The pipeline is also available as a series of scripts runnable locally through Python 3.8 with several well-maintained programs as dependencies, including Bowtie 2, Samtools, Picard, and Picard extension Drop-seq Tools *(15, 22–24)*

**Input and reference file formats**

The required input files to TCRGO include the TCR sequencing data and a reference FASTA file. Example reference FASTA files for human (*Homo sapiens*) and mouse (*Mus musculus*) are available (https://github.com/ShalekLab/tcrgo/tree/master/references). These reference files are multiple sequence FASTA files, which include a separate sequence for each known TCR V and J region. For each region, the description line is written as '>TR[AB][VJC][1-9-]+*' with the following line containing the sequence.

TCR sequencing data may be represented as a single-ended FASTQ file of the 150 bp TCR read and the Seq-Well S$^3$ bead barcode and UMI represented as the index sequence portion of the sequence description (Fig. 7B). Alternatively, the TCR sequencing data may be represented as paired-end FASTQ files with the TCR read on one end and the Seq-Well S$^3$ bead barcode and UMI on the other end (Fig. 7A). When files of this format are input into the TCRGO pipeline, they should be specified with the barcode FASTQ as the first input FASTQ and the TCR FASTQ as the second input FASTQ.

**Fig. 7** Examples of FASTQ file structure for the two types of libraries. **A)** Files output from a run which includes an Illumina barcode after demultiplexing with the 'create-fastq-for-index-reads' option. The read 2 FASTQ contains the reverse complement of the UMI and bead barcode while the Read 1 FASTQ contains the reverse complement of the TCR sequence which can be mapped to the V, J, and C regions and the CDR3 can be reconstructed. The Illumina barcode serves to identify the original library of origin of each read in the FASTQ. **B)** Shows the file

output from a run which does not include Illumina barcodes. This file contains the same information as the barcoded files in **A** but relies on the bead barcode to identify the library of origin of the TCR read.

The TCRGO pipeline is highly tunable to user input specifications and allows a variety of input arguments. Several arguments are provided to adjust various thresholds. In the filter queries step, the argument 'minimum_reads', with a default of 5, adjusts the minimum threshold of number of reads assigned to a single UMI to include that UMI in the following alignment steps. During the CDR3 recovery step, several thresholds of interest control the outputs. The 'minimum_frequency' threshold describes, for a given UMI, the minimum frequency of total reads that must be aligned to the top V/J combination for this UMI to be included. This threshold mostly filters out UMIs with few reads, especially when those reads map to very different places. Relatedly, the 'minimum_cdr3s' threshold determines the minimum number of reads mapped to the top V/J combination necessary to proceed with CDR3 assembly for a given UMI. If the library has generally lower read depth and results in lower total reads, another parameter to tune to increase the number of recovered TCRs without relaxing this threshold is the 'exclude_relatives' parameter. When selected, this parameter directs the program to use reads mapped to any sequence in the TCR family of the top V/J regions in assembling the CDR3. For example, if a subset of reads is mapped to TRAV4-1 and another subset to TRAV4-2, reads mapped to both of these regions would be used to construct the CDR3 sequence. Further input arguments include technical specifications of disc sizes as well as parameters to tune the extent of parallelization of the pipeline. In depth descriptions of these are available at

https://github.com/ShalekLab/tcrgo#list-of-inputs.

**Preprocessing and alignment**

The preprocessing and alignment step formats input files, performs quality control, and aligns the reads to the reference. It prepares the input files for alignment by reformatting them to single-end BAM format to allow the use of the Drop-seq Tools pipeline for quality control. Using components of the Drop-seq Tools program, the reads are tagged with cell and molecular barcodes, low quality barcodes are filtered, and primer and polyA sequences are trimmed. Multimap alignment to V, J, and C TRA and TRB gene segments is performed via Bowtie 2 while all other preprocessing tasks are handled by Samtools, Picard, and Picard extension Drop-seq Tools *(15, 22–24)*. We expect a very high mapping percentage (around 98%) and that a majority of reads map to multiple locations because each read should map to at least a V and J region (Fig. 7A).

**Filtering**

The filtering step acts to remove reads which do not contain both V and J mapping information. It iterates over the alignment data and distributes reads that multimap to both a V and J segment to lists which can be used in parallel by the CDR3 recovery step.

**CDR3 Recovery**

The CDR3 recovery step quantifies the alignment identity data for each read, then it resolves the V and J identities for each UMI and determines the UMI's most likely CDR3 sequence using a greedy, optimized, heuristic-driven algorithm.

First, the V and J regions are identified for each read. Often, a single read maps to multiple V or J regions, so the top V and J alignments are determined by Bowtie 2 alignment score, or 'AS' tag. If there is a tie in the alignment score, several metrics are considered. First, the edit distances to the reference sequences are considered. If the tie continues, the Phred quality of the alignment is considered. Then, if the alignments span different lengths of the read, the longer alignment is used. If all options result in a tie, the top alignment is determined alphabetically. If a tie occurs, the information about the tie and the method used to break the tie is recorded for user debugging. Each UMI is assigned the V and J regions to which the most reads were mapped. Users may specify a minimum frequency threshold to filter for UMIs for which a minimum proportion of reads map to their most frequent V and J regions. This threshold allows for more stringent filtering of UMIs without a conclusive mapping location due to sequencing errors or other technical artifacts.

**CDR3 positions file**

The --cdr3-positions-file argument can be used to input a 2-column TSV file containing the names of FASTA TR(A|B)(V|J) segments and that entry's corresponding CDR3 start or ending codon position in the nucleotide sequence. Each entry name in the CDR3 positions file must exactly match its respective entry in the FASTA. If this file is provided, the CDR3 sequences are identified using the information about the mapped V and J regions and the locations in this file. Since small sequencing errors or annotation differences may shift these locations by one base pair, the amino acid sequence of these regions is reported.

If a CDR3 positions file is not included, a greedy algorithm is used to identify this sequence for each read which maps to the most likely V and J regions. Starting at the V alignment ending position and ending at the J alignment starting position, potential CDR3 sequences are scored based on the absence of internal stop codons, starting codon encoding amino acid cysteine and ending codon encoding the amino acid phenylalanine as these are the conserved starting and stopping codons of the CDR3 region.

To resolve the CDR3 sequence from multiple reads with slightly different sequences, the most frequent CDR3 sequence is used. If there is a tie for the most frequent sequence, a consensus sequence is greedily built from the tied sequences. Ties are found most frequently for UMIs with low numbers of reads.

**Output format**

The alignment identity and CDR3 information is gathered and tabulated during the summary step as a tab-separated value text file. Each unique UMI and barcode pair is output as a single line in the summary table which includes the read counts for that UMI, V and J alignment information, and CDR3 consensus sequence, example available at [https://github.com/ShalekLab/tcrgo/tree/master/examples](https://github.com/ShalekLab/tcrgo/tree/master/examples) (Table S1). If the 'collapse' argument is specified, barcode/UMI pairs are collapsed based on similarity of V/J mapping and CDR3 sequence and an additional output file ending with 'aggrcollapsed_cdr3_info.tsv' is generated.

### 3.5.3. Mapping to Seq-Well S$^3$ cells

After reads are assigned to cell barcodes and UMIs, these TCRs must be assigned to T cells found in the full Seq-Well S$^3$ transcriptomic data. To do this, we obtain a list of cell barcode sequences from the Seq-Well S$^3$ data matrix and identify the most similar by Hamming distance Seq-Well S$^3$ barcode to each TCR barcode. Example python code for this application is available at https://github.com/ShalekLab/tcrgo/tree/master/examples.

4. **Notes**

1. V region primer design requires access to an annotation of known V region sequences for the species of interest such as those available in IMGT *(25)*. These primer mixes are designed manually such that each V region may be bound by a maximum of one primer to prevent competitive binding. The binding region is preferred to be found near the middle of the V sequence to allow sufficient sequence length to uniquely identify that region during alignment and allow the sequencing read to span the CDR3 region. These sequences are approximately 60 bases long and have a melting point near 70 ºC. The first 30 bases of the sequence include the UPS2 sequence, and the remaining bases of the sequence are complementary to a V region. Primer sequences have been optimized for human and mouse (Table S5B, C) and are undergoing testing for crab-eating macaques (Table S5A).

2. This protocol has been optimized for the following TCR probes which are complementary to the alpha and beta constant regions of the corresponding species (Table S2). When multiple probes for a single constant region are listed, they bind to two overlapping regions of the constant region and should be used if sequential enrichment steps are run.

3. There are two options for demultiplexing TCR regions from pooled sequencing libraries. The first simply uses the Seq-Well S$^3$ bead barcodes to match TCRs to their corresponding cells. If profiling a limited number of T cells, it is unlikely that you will have overlapping Seq-Well S$^3$ bead barcodes between two different T cells; thus, in certain instances, you can pool a handful of TCR libraries together into a single sequencing runs without additional indexing barcodes and still map each read to the original Seq-Well S$^3$ array and T cell of origin. Alternatively, each TCR sample can be indexed by adding a barcode into the UPS2-N50x primers (Table S3). If this is done, primers with different index sequences should be used for each sample and no samples tagged with sequences which differ by 2 bp or fewer should be added to the same sequencing lane.

4. This protocol has been optimized for the following custom read 1 sequencing primers (Table S4). These sequencing primers are complementary to the constant region of the TCR alpha or beta and are species-specific unless otherwise noted.

5. Thaw all buffers at RT, except 10X Wash buffer 1 which may need to be heated at 37ºC to ensure all salts go into solution and to resuspend particulates.

6. If using a thermocycler, the lid should be set to 105 ºC.

7. Ensure beads are resuspended and that they do not dry in the well before the addition of a sample. Small amounts of Bead Wash Buffer will not interfere with downstream binding of the DNA to the beads.

8. Here we recommend using a thermal shaker to facilitate intermittent vortex.

9. The buffers should be heated at 65 ºC for at least 15 min before continuing to wash steps.

10. To minimize non-specific binding of off-target DNA sequences to the capture probes, work quickly to prevent the temperature of the hybridized samples from dropping significantly at temperature sensitive steps.

11. 0.6 means the ratio of AMPure XP reagent volume to sample volume. For example, 0.6 of AMPure XP reagent concentration means 60 μl AMPure XP reagent in 100 μl sample (100 x 0.6). Ratio of AMPure XP can be adjusted for improved recovery.

12. Beads should appear matte and have a slight cracked appearance. Prevent over-drying of beads as this can lead to significant yield loss.

13. Certain samples may need to be enriched multiple times, especially samples with low gene recovery or low number of α and β T cells. If the amplification was unsuccessful, we recommend decreasing the number of cycles of PCR amplification from 25 to 12 for subsequent rounds of enrichment.

14. Slight variation between samples is acceptable and expected. After affinity purification, we see peaks that range between 1-2 kB. Upon completing the full pulldown, we expect TCRα peaks around ~1.3 kB and TCRβ peaks ~1.1 kB. DNA contamination from miscellaneous products from WTA are expected and may result in traces with multiple peaks. However, these products will not be sequenced.

15. This protocol has been optimized for the following TCR V-region primers (Table S5A-C).

16. Here to minimize barcode swapping, we split into 4 PCR reactions. Only the UPS-500 primers will have a barcode. You may choose to rely on Seq-Well S[3] bead barcodes for demultiplexing and forgo the Illumina barcode.

17. The precise number of cycles needs to be tested and adjusted for each sample. **Table 8** is a recommendation based on the concentration estimation of amplified pull-down products from **Section 3.1.8**.

18. If amplification was unsuccessful after AMPure XP purification, the remaining half of the amplified product can be further amplified for 2-4 PCR cycles.

19. It is expected that the concentration of samples obtained from the fragment analyzer will be higher than the qPCR determined concentration due to the presence of miscellaneous DNA products from the original WTA.

20. For libraries consisting of a high number of cells, a NextSeq could be used. For NextSeq, sequencing primers are used at a final concentration of 2.5 µM. We aimed for $150\text{-}200 \times 10^6$ pass filter reads per lane, a cluster density of roughly $100{,}000/\text{mm}^2$). Based on the whole-transcriptome data, we allotted ~80,000 T cells per lane.

21. Due to the large fragment length of the final libraries, we aim to load the flow cells at a lower cluster density.

22. For sequencing with Illumina index barcodes, Read 1 should be assigned as 140 bp, Index 1 should be 20 bp, and Index 2 should be assigned as 8 bp.

**Supplementary Table Captions**

**Supplementary Table 1** Example output of TCRGO pipeline.

**Supplementary Table 2** Biotinylated oligonucleotide probes for TCR enrichment for human (*Homo sapiens*), mouse (*Mus musculus*), and crab-eating macaque (*Macaca fasciculari*s, abbreviated MacFas), respectively.

**Supplementary Table 3** UPS2_N50x primers with Illumina barcodes.

**Supplementary Table 4** Optimized custom TCR sequencing primers for human, mouse and crab-eating macaque. Human/macaque primers are universal for human and macaque species.

**Supplementary Tables 5A-5C** Species specific variable region primers (A) human (B) mouse, (C) crab-eating macaque (MacFas).

Supplementary_Table1_Example_TCRGO_output

| BC_index | UMI_index | BC | UMI | nReads | topN2_region | topN2_nReads | topN1_freq | CDR3_nt | CDR3_aa | CDR3_isProductive | CDR3_nReads | CDR3_freq | CDR3_stopcodons |
|---|---|---|---|---|---|---|---|---|---|---|---|---|---|

| TRAV_nReads | TRAJ_nReads | TRAC_nReads | TRBV_nReads | TRBJ_nReads | TRBC_nReads | UNKN_nReads |
|---|---|---|---|---|---|---|
| 23 | 23 | 0 | 0 | 0 | 0 | 0 |
| 16 | 16 | 0 | 0 | 0 | 0 | 0 |
| 7 | 7 | 0 | 0 | 0 | 0 | 0 |
| 19 | 19 | 0 | 0 | 0 | 0 | 0 |
| 23 | 23 | 0 | 0 | 0 | 0 | 0 |
| 80 | 80 | 0 | 0 | 0 | 0 | 0 |
| 1000 | 1000 | 0 | 0 | 0 | 0 | 0 |
| 10 | 10 | 0 | 0 | 0 | 0 | 0 |
| 8 | 8 | 0 | 0 | 0 | 0 | 0 |
| 29 | 29 | 0 | 0 | 0 | 0 | 0 |
| 56 | 56 | 0 | 0 | 0 | 0 | 0 |
| 58 | 58 | 0 | 0 | 0 | 0 | 0 |
| 8 | 8 | 0 | 0 | 0 | 0 | 0 |
| 50 | 50 | 0 | 0 | 0 | 0 | 0 |
| 14 | 14 | 0 | 0 | 0 | 0 | 0 |
| 45 | 45 | 0 | 0 | 0 | 0 | 0 |
| 0 | 0 | 0 | 5 | 5 | 0 | 0 |
| 7 | 7 | 0 | 0 | 0 | 0 | 0 |
| 15 | 15 | 0 | 0 | 0 | 0 | 0 |
| 14 | 14 | 0 | 0 | 0 | 0 | 0 |
| 17 | 17 | 0 | 0 | 0 | 0 | 0 |
| 13 | 13 | 0 | 0 | 0 | 0 | 0 |
| 140 | 140 | 0 | 0 | 0 | 0 | 2 |
| 8 | 8 | 0 | 0 | 0 | 0 | 0 |
| 0 | 0 | 0 | 8 | 8 | 0 | 0 |
| 0 | 0 | 0 | 18 | 18 | 0 | 0 |
| 26 | 26 | 0 | 0 | 0 | 0 | 0 |
| 8 | 8 | 0 | 0 | 0 | 0 | 0 |
| 23 | 23 | 0 | 0 | 0 | 0 | 0 |
| 8 | 8 | 0 | 0 | 0 | 0 | 0 |
| 97 | 97 | 0 | 0 | 0 | 0 | 0 |
| 7 | 7 | 0 | 0 | 0 | 0 | 0 |
| 137 | 137 | 0 | 0 | 0 | 0 | 0 |
| 31 | 31 | 0 | 0 | 0 | 0 | 0 |
| 139 | 139 | 0 | 0 | 0 | 0 | 0 |
| 23 | 23 | 0 | 0 | 0 | 0 | 0 |
| 7 | 7 | 0 | 0 | 0 | 0 | 0 |
| 63 | 63 | 0 | 0 | 0 | 0 | 0 |
| 44 | 44 | 0 | 0 | 0 | 0 | 0 |
| 47 | 47 | 0 | 0 | 0 | 0 | 0 |
| 6 | 6 | 0 | 0 | 0 | 0 | 0 |
| 12 | 12 | 0 | 0 | 0 | 0 | 0 |
| 6 | 6 | 0 | 0 | 0 | 0 | 0 |
| 7 | 7 | 0 | 0 | 0 | 0 | 0 |
| 7 | 7 | 0 | 0 | 0 | 0 | 0 |
| 5 | 5 | 0 | 0 | 0 | 0 | 0 |
| 5 | 5 | 0 | 0 | 0 | 0 | 0 |
| 7 | 7 | 0 | 0 | 0 | 0 | 0 |
| 74 | 74 | 0 | 0 | 0 | 0 | 0 |
| 1000 | 1000 | 0 | 0 | 0 | 0 | 0 |
| 8 | 8 | 0 | 0 | 0 | 0 | 0 |
| 12 | 12 | 0 | 0 | 0 | 0 | 0 |
| 7 | 7 | 0 | 0 | 0 | 0 | 0 |
| 10 | 10 | 0 | 0 | 0 | 0 | 0 |
| 5 | 5 | 0 | 0 | 0 | 0 | 0 |
| 10 | 10 | 0 | 0 | 0 | 0 | 0 |
| 6 | 6 | 0 | 0 | 0 | 0 | 0 |
| 7 | 7 | 0 | 0 | 0 | 0 | 0 |
| 12 | 12 | 0 | 0 | 0 | 0 | 0 |
| 335 | 335 | 0 | 0 | 0 | 0 | 0 |
| 15 | 15 | 0 | 0 | 0 | 0 | 0 |
| 87 | 87 | 0 | 0 | 0 | 0 | 0 |
| 334 | 334 | 0 | 0 | 0 | 0 | 0 |
| 6 | 6 | 0 | 0 | 0 | 0 | 0 |
| 31 | 31 | 0 | 0 | 0 | 0 | 0 |
| 7 | 7 | 0 | 0 | 0 | 0 | 0 |
| 76 | 76 | 0 | 0 | 0 | 0 | 0 |
| 5 | 5 | 0 | 0 | 0 | 0 | 0 |
| 16 | 16 | 0 | 0 | 0 | 0 | 0 |
| 175 | 175 | 0 | 0 | 0 | 0 | 0 |
| 16 | 16 | 0 | 0 | 0 | 0 | 0 |
| 35 | 35 | 0 | 0 | 0 | 0 | 0 |
| 86 | 86 | 0 | 0 | 0 | 0 | 0 |
| 7 | 7 | 0 | 0 | 0 | 0 | 0 |
| 147 | 147 | 0 | 0 | 0 | 0 | 0 |
| 7 | 7 | 0 | 0 | 0 | 0 | 0 |
| 40 | 40 | 0 | 0 | 0 | 0 | 0 |
| 64 | 64 | 0 | 0 | 0 | 0 | 0 |
| 58 | 58 | 0 | 0 | 0 | 0 | 0 |
| 13 | 13 | 0 | 0 | 0 | 0 | 0 |
| 37 | 37 | 0 | 0 | 0 | 0 | 0 |
| 24 | 24 | 0 | 0 | 0 | 0 | 0 |
| 86 | 86 | 0 | 0 | 0 | 0 | 0 |
| 50 | 50 | 0 | 0 | 0 | 0 | 0 |
| 7 | 7 | 0 | 0 | 0 | 0 | 0 |
| 146 | 146 | 0 | 0 | 0 | 0 | 0 |
| 47 | 47 | 0 | 0 | 0 | 0 | 0 |
| 25 | 25 | 0 | 0 | 0 | 0 | 0 |

Supplementary_Table2_TCRC_Probes

| Name | Sequence | Vendor/Service |
|---|---|---|
| **Human TRBC-1** | /5BiosG/GTGTTCCCACCCRAGGTCGCTGTGTTTGAGCCATCAGAAGCAGAGATCTCCCACACCCAAAAGGCCACACTGGTGTGCCTGGCCACAGGC | IDT Ultramer/Standard Desalting |
| **Human TRAC-1** | /5BiosG/CTGTCTGCCTATTCACCCGATTTTGATTCTCAAACAAATGTGTCACAAAGTAAGGATTCTGATGTGTATATCACAGACAAAACTGTGCTAG | IDT Ultramer/Standard Desalting |
| **Mouse TRBC-1** | /5BiosG/AGGATCTGAGAAATGTGACTCCACCCAAGGTCTCCTTGTTTGAGCCATCAAAAGCAGAGATTGCAAACAAACAAAAGGCTACCCTCGTGT | IDT Ultramer/Standard Desalting |
| **Mouse TRAC-1** | /5BiosG/ACATCCAGAACCCAGAACCTGCTGTGTACCAGTTAAAGATCCTCGGTCTCAGGACAGCACCCTCTGCCTGTTCACCGACTTTGACTCCC | IDT Ultramer/Standard Desalting |
| **Macaca fascicularis (MacFas) TRBC-1** | /5BiosG/GAGGACCTGAAAAAGGTGTTCCCACCCAAGGTTGCTGTRTTTGAGCCATCAGAAGCAGAGATCTCCCACACCCAAAAGGCCACGCTGGTG | IDT Ultramer/Standard Desalting |
| **MacFas TRBC-2** | /5BiosG/GCCATCAGAAGCAGAGATCTCCCACACCCAAAAGGCCACGCTGGTGTGCCTGGCCACAGGCTTCTACCCCGACCACGTGGAGCTGAGCTG | IDT Ultramer/Standard Desalting |
| **MacFas TRAC-1** | /5BiosG/ATATCCAGAACCCTGACCCTGCCGTGTACCAGCTGAGAGGCTCTAAATCCAATGACACCTCTGTCTGCCTATTTACTGATTTTGATTTCTG | IDT Ultramer/Standard Desalting |
| **MacFas TRAC-2** | /5BiosG/TAAATCCAATGACACCTCTGTCTGCCTATTTACTGATTTTGATTCTGTAATGAATGTGTCACAAAGCAAGGATTCTGACGTGCATATCAC | IDT Ultramer/Standard Desalting |

Supplementary_Table3_UPS2_N50x_indexed_primers

| Well Position | Name | Sequence |
|---|---|---|
| A1 | P5-TSO_Hybrid_N501 | AATGATACGGCGACCACCGAGATCTACACTAGATCGCGCCTGTCCGCGGAAGCAGTGGTATCAACGCAGAGT*A*C |
| B1 | P5-TSO_Hybrid_N502 | AATGATACGGCGACCACCGAGATCTACACCTCTCTATGCCTGTCCGCGGAAGCAGTGGTATCAACGCAGAGT*A*C |
| C1 | P5-TSO_Hybrid_N503 | AATGATACGGCGACCACCGAGATCTACACTATCCTCTGCCTGTCCGCGGAAGCAGTGGTATCAACGCAGAGT*A*C |
| D1 | P5-TSO_Hybrid_N504 | AATGATACGGCGACCACCGAGATCTACACAGAGTAGAGCCTGTCCGCGGAAGCAGTGGTATCAACGCAGAGT*A*C |
| E1 | P5-TSO_Hybrid_N505 | AATGATACGGCGACCACCGAGATCTACACGTAAGGAGGCCTGTCCGCGGAAGCAGTGGTATCAACGCAGAGT*A*C |
| F1 | P5-TSO_Hybrid_N506 | AATGATACGGCGACCACCGAGATCTACACACTGCATAGCCTGTCCGCGGAAGCAGTGGTATCAACGCAGAGT*A*C |
| G1 | P5-TSO_Hybrid_N507 | AATGATACGGCGACCACCGAGATCTACACAAGGAGTAGCCTGTCCGCGGAAGCAGTGGTATCAACGCAGAGT*A*C |
| H1 | P5-TSO_Hybrid_N508 | AATGATACGGCGACCACCGAGATCTACACCTAAGCCTGCCTGTCCGCGGAAGCAGTGGTATCAACGCAGAGT*A*C |
| A2 | P5-TSO_Hybrid_N517 | AATGATACGGCGACCACCGAGATCTACACGCGTAAGAGCCTGTCCGCGGAAGCAGTGGTATCAACGCAGAGT*A*C |
| B2 | P5-TSO_Hybrid_N510 | AATGATACGGCGACCACCGAGATCTACACCGTCTAATGCCTGTCCGCGGAAGCAGTGGTATCAACGCAGAGT*A*C |
| C2 | P5-TSO_Hybrid_N511 | AATGATACGGCGACCACCGAGATCTACACTCTCTCCGGCCTGTCCGCGGAAGCAGTGGTATCAACGCAGAGT*A*C |
| D2 | P5-TSO_Hybrid_N513 | AATGATACGGCGACCACCGAGATCTACACTCGACTAGGCCTGTCCGCGGAAGCAGTGGTATCAACGCAGAGT*A*C |
| E2 | P5-TSO_Hybrid_N515 | AATGATACGGCGACCACCGAGATCTACACTTCTAGCTGCCTGTCCGCGGAAGCAGTGGTATCAACGCAGAGT*A*C |
| F2 | P5-TSO_Hybrid_N516 | AATGATACGGCGACCACCGAGATCTACACCCTAGAGTGCCTGTCCGCGGAAGCAGTGGTATCAACGCAGAGT*A*C |
| G2 | P5-TSO_Hybrid_N518 | AATGATACGGCGACCACCGAGATCTACACCTATTAAGGCCTGTCCGCGGAAGCAGTGGTATCAACGCAGAGT*A*C |
| H2 | P5-TSO_Hybrid_N520 | AATGATACGGCGACCACCGAGATCTACACAAGGCTATGCCTGTCCGCGGAAGCAGTGGTATCAACGCAGAGT*A*C |
| A3 | P5-TSO_Hybrid_N522 | AATGATACGGCGACCACCGAGATCTACACTTATGCGAGCCTGTCCGCGGAAGCAGTGGTATCAACGCAGAGT*A*C |
| B3 | P5-TSO_Hybrid_N521 | AATGATACGGCGACCACCGAGATCTACACGAGCCTTAGCCTGTCCGCGGAAGCAGTGGTATCAACGCAGAGT*A*C |
| C3 | P5-TSO_Hybrid_N5_BC01 | AATGATACGGCGACCACCGAGATCTACACAACAATGGGCCTGTCCGCGGAAGCAGTGGTATCAACGCAGAGT*A*C |
| E3 | P5-TSO_Hybrid_N5_BC03 | AATGATACGGCGACCACCGAGATCTACACAACTTGACGCCTGTCCGCGGAAGCAGTGGTATCAACGCAGAGT*A*C |
| F3 | P5-TSO_Hybrid_N5_BC04 | AATGATACGGCGACCACCGAGATCTACACTTCGCTGAGCCTGTCCGCGGAAGCAGTGGTATCAACGCAGAGT*A*C |
| H3 | P5-TSO_Hybrid_N5_BC06 | AATGATACGGCGACCACCGAGATCTACACTTTACGCACGCCTGTCCGCGGAAGCAGTGGTATCAACGCAGAGT*A*C |
| A4 | P5-TSO_Hybrid_N5_BC07 | AATGATACGGCGACCACCGAGATCTACACAATGTTCTGCCTGTCCGCGGAAGCAGTGGTATCAACGCAGAGT*A*C |
| B4 | P5-TSO_Hybrid_N5_BC08 | AATGATACGGCGACCACCGAGATCTACACTGTCGGATGCCTGTCCGCGGAAGCAGTGGTATCAACGCAGAGT*A*C |
| C4 | P5-TSO_Hybrid_N5_BC09 | AATGATACGGCGACCACCGAGATCTACACACAGGTATGCCTGTCCGCGGAAGCAGTGGTATCAACGCAGAGT*A*C |
| D4 | P5-TSO_Hybrid_N5_BC10 | AATGATACGGCGACCACCGAGATCTACACGTGTAACTCGCCTGTCCGCGGAAGCAGTGGTATCAACGCAGAGT*A*C |
| E4 | P5-TSO_Hybrid_N5_BC11 | AATGATACGGCGACCACCGAGATCTACACCACCAACTGGCCTGTCCGCGGAAGCAGTGGTATCAACGCAGAGT*A*C |
| F4 | P5-TSO_Hybrid_N5_BC12 | AATGATACGGCGACCACCGAGATCTACACTGCTCGACGCCTGTCCGCGGAAGCAGTGGTATCAACGCAGAGT*A*C |
| H4 | P5-TSO_Hybrid_N5_BC14 | AATGATACGGCGACCACCGAGATCTACACTCTGGCGAGCCTGTCCGCGGAAGCAGTGGTATCAACGCAGAGT*A*C |
| A5 | P5-TSO_Hybrid_N5_BC15 | AATGATACGGCGACCACCGAGATCTACACAGCATGGACCTGTCCGCGGAAGCAGTGGTATCAACGCAGAGT*A*C |
| B5 | P5-TSO_Hybrid_N5_BC16 | AATGATACGGCGACCACCGAGATCTACACTCTCGGTCGCCTGTCCGCGGAAGCAGTGGTATCAACGCAGAGT*A*C |
| C5 | P5-TSO_Hybrid_N5_BC17 | AATGATACGGCGACCACCGAGATCTACACAGGTAAGGGCCTGTCCGCGGAAGCAGTGGTATCAACGCAGAGT*A*C |
| D5 | P5-TSO_Hybrid_N5_BC18 | AATGATACGGCGACCACCGAGATCTACACTCGCTAGAGCCTGTCCGCGGAAGCAGTGGTATCAACGCAGAGT*A*C |
| E5 | P5-TSO_Hybrid_N5_BC19 | AATGATACGGCGACCACCGAGATCTACACAGGTGCGAGCCTGTCCGCGGAAGCAGTGGTATCAACGCAGAGT*A*C |
| F5 | P5-TSO_Hybrid_N5_BC20 | AATGATACGGCGACCACCGAGATCTACACTCCTTGGTGCCTGTCCGCGGAAGCAGTGGTATCAACGCAGAGT*A*C |
| G5 | P5-TSO_Hybrid_N5_BC21 | AATGATACGGCGACCACCGAGATCTACACAGTTGCTTGCCTGTCCGCGGAAGCAGTGGTATCAACGCAGAGT*A*C |
| H5 | P5-TSO_Hybrid_N5_BC22 | AATGATACGGCGACCACCGAGATCTACACTATCTGCCGCCTGTCCGCGGAAGCAGTGGTATCAACGCAGAGT*A*C |
| A6 | P5-TSO_Hybrid_N5_BC23 | AATGATACGGCGACCACCGAGATCTACACATTATCAAGCCTGTCCGCGGAAGCAGTGGTATCAACGCAGAGT*A*C |
| B6 | P5-TSO_Hybrid_N5_BC24 | AATGATACGGCGACCACCGAGATCTACACTACTTAGCGCCTGTCCGCGGAAGCAGTGGTATCAACGCAGAGT*A*C |
| D6 | P5-TSO_Hybrid_N5_BC26 | AATGATACGGCGACCACCGAGATCTACACTAAGCACAGCCTGTCCGCGGAAGCAGTGGTATCAACGCAGAGT*A*C |
| E6 | P5-TSO_Hybrid_N5_BC27 | AATGATACGGCGACCACCGAGATCTACACATTGTCTGGCCTGTCCGCGGAAGCAGTGGTATCAACGCAGAGT*A*C |
| F6 | P5-TSO_Hybrid_N5_BC28 | AATGATACGGCGACCACCGAGATCTACACGTCGCCTTGCCTGTCCGCGGAAGCAGTGGTATCAACGCAGAGT*A*C |
| G6 | P5-TSO_Hybrid_N5_BC29 | AATGATACGGCGACCACCGAGATCTACACCAATAGTGCGCCTGTCCGCGGAAGCAGTGGTATCAACGCAGAGT*A*C |

| H6 | P5-TSO_Hybrid_N5_BC30 | AATGATACGGCGACCACCGAGATCTACACGTCATCTAGCCTGTCCGCGGAAGCAGTGGTATCAACGCAGAGT*A*C |
| A7 | P5-TSO_Hybrid_N5_BC31 | AATGATACGGCGACCACCGAGATCTACACCAGCAAGGGCCTGTCCGCGGAAGCAGTGGTATCAACGCAGAGT*A*C |
| B7 | P5-TSO_Hybrid_N5_BC32 | AATGATACGGCGACCACCGAGATCTACACGTAACATCGCCTGTCCGCGGAAGCAGTGGTATCAACGCAGAGT*A*C |
| C7 | P5-TSO_Hybrid_N5_BC33 | AATGATACGGCGACCACCGAGATCTACACCAGGAGCCGCCTGTCCGCGGAAGCAGTGGTATCAACGCAGAGT*A*C |
| D7 | P5-TSO_Hybrid_N5_BC34 | AATGATACGGCGACCACCGAGATCTACACGCCTAGCCGCCTGTCCGCGGAAGCAGTGGTATCAACGCAGAGT*A*C |
| E7 | P5-TSO_Hybrid_N5_BC35 | AATGATACGGCGACCACCGAGATCTACACCATGATCGGCCTGTCCGCGGAAGCAGTGGTATCAACGCAGAGT*A*C |
| F7 | P5-TSO_Hybrid_N5_BC36 | AATGATACGGCGACCACCGAGATCTACACGCCGCAACGCCTGTCCGCGGAAGCAGTGGTATCAACGCAGAGT*A*C |
| G7 | P5-TSO_Hybrid_N5_BC37 | AATGATACGGCGACCACCGAGATCTACACCCAACATTGCCTGTCCGCGGAAGCAGTGGTATCAACGCAGAGT*A*C |
| H7 | P5-TSO_Hybrid_N5_BC38 | AATGATACGGCGACCACCGAGATCTACACGCACATCTGCCTGTCCGCGGAAGCAGTGGTATCAACGCAGAGT*A*C |

Supplementary_Table4_TCR_SequencingPrimers

| Name | Sequence | Vendor/Service |
|---|---|---|
| **Human Alpha Sequencing Primer** | AGAGTCTCTCAGCTGGTACACGGCAGGGTCAGGITTCTGGATAT | IDT/HPLC |
| **Human Beta Sequencing Primer** | CAAACACAGCGACCTCGGGTGGGAACACSTTKTTCAGGTCCT | IDT/HPLC |
| **Mouse Alpha Sequencing Primer** | GTCCTGAGACCGAGGATCTTTTAACTGGTACACAGCAGGTTCTGGGTTCTGGATGT | IDT/HPLC |
| **Mouse Beta Sequencing Primer** | TGCTTTTGATGGCTCAAACAAGGAGACCTTGGGTGGAGTCACATTTCTCAGATCCT | IDT/HPLC |
| **Human/Macaque Alpha Sequencing Primer** | AGAGICTCTCAGCTGGTACACGGCAGGGTCAGGITTCTGGATAT | IDT/HPLC |
| **Human/Macaque Beta Sequencing Primer** | CAAACACAGCIACCTIGGGTGGGAACACSTTKTTCAGGTCCT | IDT/HPLC |

## Supplementary_Table5A_Human_TCRV_Primers

| Sequence Name | Sequence |
|---|---|
| **hTRAV1.UPS2** | TCG TGG GCT CGG AGA TGT GTA TAA GAG ACA GAG GTC GTT TTT CTT CAT TCC TTA GTC |
| **hTRAV2.UPS2** | TCG TGG GCT CGG AGA TGT GTA TAA GAG ACA GAC GAT ACA ACA TGA CCT ATG AAC GG |
| **hTRAV3-1.UPS2** | TCG TGG GCT CGG AGA TGT GTA TAA GAG ACA GCT TTG AAG CTG AAT TTA ACA AGA GCC |
| **hTRAV4-1.UPS2** | TCG TGG GCT CGG AGA TGT GTA TAA GAG ACA GCT CCC TGT TTA TCC CTG CCG AC |
| **hTRAV5-1.UPS2** | TCG TGG GCT CGG AGA TGT GTA TAA GAG ACA GAA ACA AGA CCA AAG ACT CAC TGT TC |
| **hTRAV6.UPS2** | TCG TGG GCT CGG AGA TGT GTA TAA GAG ACA GAA GAC TGA AGG TCA CCT TTG ATA CC |
| **hTRAV7.UPS2** | TCG TGG GCT CGG AGA TGT GTA TAA GAG ACA GAC TAA ATG CTA CAT TAC TGA AGA ATG G |
| **hTRAV8.UPS2** | TCG TGG GCT CGG AGA TGT GTA TAA GAG ACA GGC ATC AAC GGT TTT GAG GCT GAA TTT AA |
| **hTRAV9.UPS2** | TCG TGG GCT CGG AGA TGT GTA TAA GAG ACA GGA AAC CAC TTC TTT CCA CTT GGA GAA |
| **hTRAV10.UPS2** | TCG TGG GCT CGG AGA TGT GTA TAA GAG ACA GTA CAG CAA CTC TGG ATG CAG ACA C |
| **hTRAV12.UPS2** | TCG TGG GCT CGG AGA TGT GTA TAA GAG ACA GGA AGA TGG AAG GTT TAC AGC ACA |
| **hTRAV13-1.UPS2** | TCG TGG GCT CGG AGA TGT GTA TAA GAG ACA GGA CAT TCG TTC AAA TGT GGG CGA A |
| **hTRAV13-2.UPS2** | TCG TGG GCT CGG AGA TGT GTA TAA GAG ACA GGG CAA GGC CAA AGA GTC ACC GT |
| **hTRAV14.UPS2** | TCG TGG GCT CGG AGA TGT GTA TAA GAG ACA GTC CAG AAG GCA AGA AAA TCC GCC A |
| **hTRAV16.UPS2** | TCG TGG GCT CGG AGA TGT GTA TAA GAG ACA GGC TGA CCT TAA CAA AGG CGA GAC A |
| **hTRAV17.UPS2** | TCG TGG GCT CGG AGA TGT GTA TAA GAG ACA GTT AAG AGT CAC GCT TGA CAC TTC CA |
| **hTRAV18.UPS2** | TCG TGG GCT CGG AGA TGT GTA TAA GAG ACA GGC AGA GGT TTT CAG GCC AGT CCT |
| **hTRAV19.UPS2** | TCG TGG GCT CGG AGA TGT GTA TAA GAG ACA GTC CAC CAG TTC CTT CAA CTT CAC C |
| **hTRAV20.UPS2** | TCG TGG GCT CGG AGA TGT GTA TAA GAG ACA GGC CAC ATT AAC AAA GAA GGA AAG CT |
| **hTRAV21.UPS2** | TCG TGG GCT CGG AGA TGT GTA TAA GAG ACA GGC CTC GCT GGA TAA ATC ATC AGG A |
| **hTRAV22.UPS2** | TCG TGG GCT CGG AGA TGT GTA TAA GAG ACA GAC GAC TGT CGC TAC GGA ACG CTA |
| **hTRAV23.UPS2** | TCG TGG GCT CGG AGA TGT GTA TAA GAG ACA GCA CAA TCT CCT TCA ATA AAA GTG CCA |
| **hTRAV24.UPS2** | TCG TGG GCT CGG AGA TGT GTA TAA GAG ACA GAC GAA TAA GTG CCA CTC TTA ATA CCA |
| **hTRAV25.UPS2** | TCG TGG GCT CGG AGA TGT GTA TAA GAG ACA GGT TTG GAG AAG CAA AAA AGA ACA GCT |
| **hTRAV26-1.UPS2** | TCG TGG GCT CGG AGA TGT GTA TAA GAG ACA GCA GAA GAC AGA AAG TCC AGC ACC T |
| **hTRAV26-2.UPS2** | TCG TGG GCT CGG AGA TGT GTA TAA GAG ACA GAT CGC TGA AGA CAG AAA GTC CAG T |
| **hTRAV27.UPS2** | TCG TGG GCT CGG AGA TGT GTA TAA GAG ACA GAC TAA CCT TTC AGT TTG GTG ATG CAA |
| **hTRAV29.UPS2** | TCG TGG GCT CGG AGA TGT GTA TAA GAG ACA GCT TAA ACA AAA GTG CCA AGC ACC TC |
| **hTRAV30.UPS2** | TCG TGG GCT CGG AGA TGT GTA TAA GAG ACA GAA TAT CTG CTT CAT TTA TGA AAA AGC |
| **hTRAV34.UPS2** | TCG TGG GCT CGG AGA TGT GTA TAA GAG ACA GCC AAG TTG GAT GAG AAA AAG CAG CA |
| **hTRAV35.UPS2** | TCG TGG GCT CGG AGA TGT GTA TAA GAG ACA GCT CAG TTT GGT ATA ACC AGA AAG GA |
| **hTRAV36.UPS2** | TCG TGG GCT CGG AGA TGT GTA TAA GAG ACA GGG AAG ACT AAG TAG CAT ATT AGA TAA G |
| **hTRAV38.UPS2** | TCG TGG GCT CGG AGA TGT GTA TAA GAG ACA GCT GTG AAC TTC CAG AAA GCA GCC A |
| **hTRAV39.UPS2** | TCG TGG GCT CGG AGA TGT GTA TAA GAG ACA GCC TCA CTT GAT ACC AAA GCC CGT |
| **hTRAV40.UPS2** | TCG TGG GCT CGG AGA TGT GTA TAA GAG ACA GAG GCG GAA ATA TTA AAG ACA AAA ACT C |
| **hTRAV41.UPS2** | TCG TGG GCT CGG AGA TGT GTA TAA GAG ACA GGA TTA ATT GCC ACA ATA AAC ATA CAG G |
| **hTRBV2.UPS2** | TCG TGG GCT CGG AGA TGT GTA TAA GAG ACA GGC CTG ATG GAT CAA ATT TCA CTC TG |
| **hTRBV3-1.UPS2** | TCG TGG GCT CGG AGA TGT GTA TAA GAG ACA GTC TCA CCT AAA TCT CCA GAC AAA GCT |
| **hTRBV4.UPS2** | TCG TGG GCT CGG AGA TGT GTA TAA GAG ACA GCC TGA ATG CCC CAA CAG CTC TC |
| **hTRBVS-48.UPS2** | TCG TGG GCT CGG AGA TGT GTA TAA GAG ACA GCT CTG AGC TGA ATG TGA ACG CCT |
| **hTRBVS-1.UPS2** | TCG TGG GCT CGG AGA TGT GTA TAA GAG ACA GCG ATT CTC AGG GCG CCA GTT CTC T |
| **hTRBV6-1.UPS2** | TCG TGG GCT CGG AGA TGT GTA TAA GAG ACA GTG GCT ACA ATG TCT CCA GAT TAA ACA A |
| **hTRBV6-23.UPS2** | TCG TGG GCT CGG AGA TGT GTA TAA GAG ACA GCC CTG ATG GCT ACA ATG TCT CCA GA |
| **hTRBV6-4.UPS2** | TCG TGG GCT CGG AGA TGT GTA TAA GAG ACA GGT GTC TCC AGA GCA AAC ACA GAT GAT T |
| **hTRBV6-56.UPS2** | TCG TGG GCT CGG AGA TGT GTA TAA GAG ACA GGT CTC CAG ATC AAC CAC AGA GGA T |
| **hTRBV6-8.UPS2** | TCG TGG GCT CGG AGA TGT GTA TAA GAG ACA GGT CTC TAG ATT AAA CAC AGA GGA TTT C |
| **hTRBV6-9.UPS2** | TCG TGG GCT CGG AGA TGT GTA TAA GAG ACA GGG CTA CAA TGT ATC CAG ATC AAA CA |
| **hTRBV7-2.UPS2** | TCG TGG GCT CGG AGA TGT GTA TAA GAG ACA GTC GCT TCT CTG CAG AGA GGA CTG G |
| **hTRBV7-3.UPS2** | TCG TGG GCT CGG AGA TGT GTA TAA GAG ACA GCG GTT CTT TGC AGT CAG GCC TGA |
| **hTRBV7-8.UPS2** | TCG TGG GCT CGG AGA TGT GTA TAA GAG ACA GCC AGT GAT CGC TTC TTT GCA GAA A |
| **hTRBV?-46.UPS2** | TCG TGG GCT CGG AGA TGT GTA TAA GAG ACA GTC TCC ACT CTG AMG ATC AGC CGC A |
| **hTRBV7-7.UPS2** | TCG TGG GCT CGG AGA TGT GTA TAA GAG ACA GGC AGA GAG GCC TGA GGG ATC AT |
| **hTRBV7-9.UPS2** | TCG TGG GCT CGG AGA TGT GTA TAA GAG ACA GCT GCA GAG AGG CCT AAG GGA TCT |
| **hTRBV9.UPS2** | TCG TGG GCT CGG AGA TGT GTA TAA GAG ACA GCT CCG CAC AAC AGT TCC CTG ACT T |
| **hTRBV10-13.UPS2** | TCG TGG GCT CGG AGA TGT GTA TAA GAG ACA GCA GAT GGC TAY AGT GTC TCT AGA TCA AA |
| **hTRBV10-2.UPS2** | TCG TGG GCT CGG AGA TGT GTA TAA GAG ACA GGT TGT CTC CAG ATC CAA GAC AGA GAA |
| **hTRBV11.UPS2** | TCG TGG GCT CGG AGA TGT GTA TAA GAG ACA GGC AGA GAG GGC TCA AGG AGT AGA CT |
| **hTRBV12-34.UPS2** | TCG TGG GCT CGG AGA TGT GTA TAA GAG ACA GGC TAA GAT GCC TAA TGC ATC ATT CTC |
| **hTRBV12-5.UPS2** | TCG TGG GCT CGG AGA TGT GTA TAA GAG ACA GCT CAG CAG AGA TGC CTG ATG CAA CT |
| **hTRBV13.UPS2** | TCG TGG GCT CGG AGA TGT GTA TAA GAG ACA GTC TCA GCT CAA CAG TTC AGT GAC TA |
| **hTRBV14.UPS2** | TCG TGG GCT CGG AGA TGT GTA TAA GAG ACA GGC TGA AAG GAC TGG AGG AC GTA T |
| **hTRBV15.UPS2** | TCG TGG GCT CGG AGA TGT GTA TAA GAG ACA GGA TAA CTT CCA ATC CAG GAG GCC G |
| **hTRBV16.UPS2** | TCG TGG GCT CGG AGA TGT GTA TAA GAG ACA GGC TAA GTG CCT CCC AAA TTC ACC C |
| **hTRBV18.UPS2** | TCG TGG GCT CGG AGA TGT GTA TAA GAG ACA GGG AAC GAT TTT CTG CTG AAT TTC CCA |
| **hTRBV19.UPS2** | TCG TGG GCT CGG AGA TGT GTA TAA GAG ACA GGG TAC AGC GTC TCT CGG GAG AAG A |
| **hTRBV20-1.UPS2** | TCG TGG GCT CGG AGA TGT GTA TAA GAG ACA GGG ACA AGT TTC TCA TCA ACC ATG CAA |
| **hTRBV24-1.UPS2** | TCG TGG GCT CGG AGA TGT GTA TAA GAG ACA GTG GAT ACA GTG TCT CTC GAC AGG C |
| **hTRBV25-1.UPS2** | TCG TGG GCT CGG AGA TGT GTA TAA GAG ACA GCA ACA GTC TCC AGA ATA AGG ACG GA |
| **hTRBV27-1.UPS2** | TCG TGG GCT CGG AGA TGT GTA TAA GAG ACA GTA CAA AGT CTC TCG AAA AGA GAA GAG GA |
| **hTRBV28.UPS2** | TCG TGG GCT CGG AGA TGT GTA TAA GAG ACA GGG GGT ACA GTG TCT CTA GAG AGA |
| **hTRBV29.UPS2** | TCG TGG GCT CGG AGA TGT GTA TAA GAG ACA GGT TTC CCA TCA GCC GCC CAA ACC TA |
| **hTRBV30.UPS2** | TCG TGG GCT CGG AGA TGT GTA TAA GAG ACA GCA GAC CCC AGG ACC GGC AGT TCA T |

Supplementary_Table5B_Mouse_TCRV_Primers

| Sequence Name | Sequence |
|---|---|
| TRAV1 | ACACTCTTTCCCTACACGACGCTCTTCCGATCTAGGGAACCTTTGCTCGGGTCAAC |
| TRAV2 | ACACTCTTTCCCTACACGACGCTCTTCCGATCTGCCTATGAAGGGCAAGAAGTGAAC |
| TRAV3 | ACACTCTTTCCCTACACGACGCTCTTCCGATCTGTGGAGCAKYGCCCTCCTCAC |
| TRAV4_A for 4-1_01 4D-2_01 4-2_01 4-2_02 | ACACTCTTTCCCTACACGACGCTCTTCCGATCTTGGTCTGAGATGCAATTTTTCTACCA |
| TRAV4_B TRAV4x-3 | ACACTCTTTCCCTACACGACGCTCTTCCGATCTAACTGTGCAGTGGTTCCTACAGAAT |
| TRAV4_C TRAV4x-4 | ACACTCTTTCCCTACACGACGCTCTTCCGATCTAGAATTCCAGGGGCAGCCTCATC |
| TRAV5 74 Tm | ACACTCTTTCCCTACACGACGCTCTTCCGATCTTTTCCYTTGGTATAAGCAAGARCCTG |
| TRAV5_3P 5x-3 all pseudo | ACACTCTTTCCCTACACGACGCTCTTCCGATCTAGAGCTTTCTTTCATCCTAAGAGGC |
| TRAV6_1 6-1 6-3 6-4 | ACACTCTTTCCCTACACGACGCTCTTCCGATCTCCRACTCTTTTSTGGTATGTCCAATA |
| TRAV6_2 6-7 | ACACTCTTTCCCTACACGACGCTCTTCCGATCTTTTACGATACACTGCAACTACTCAGC |
| TRAV6_3 6-6 | ACACTCTTTCCCTACACGACGCTCTTCCGATCTTACCCTAATCTTTTCTGGTATGTTCG |
| TRAV6_4 rest | ACACTCTTTCCCTACACGACGCTCTTCCGATCTCCGGAGAAGGTCCACAGCTCCT |
| TRAV7_1 7-2 7-3 7-5 7-6 without partials | ACACTCTTTCCCTACACGACGCTCTTCCGATCTAAKGTRCAGCAGAGCCCAGAMTC |
| TRAV7_2 the rest without partials | ACACTCTTTCCCTACACGACGCTCTTCCGATCTTCGGTGGTACAGACAGCATTCTG |
| TRAV8 | ACACTCTTTCCCTACACGACGCTCTTCCGATCTGTTCAAATGAGMGAGAGAAGCGCA |
| TRAV9_1 70C tm | ACACTCTTTCCCTACACGACGCTCTTCCGATCTTATGGTGGATCCATTTACCTCTCC |
| TRAV9_2 all apart from 9-1 | ACACTCTTTCCCTACACGACGCTCTTCCGATCTGTACCCGCGGCAGGGGCTGC |
| TRAV10 | ACACTCTTTCCCTACACGACGCTCTTCCGATCTCAACTGCACTTACACAGATACTGC |
| TRAV11 | ACACTCTTTCCCTACACGACGCTCTTCCGATCTGACAACCACTTAAGGTGGTTCAAAC |
| TRAV12_1 all but 12-4_all and 12D-1_3 | ACACTCTTTCCCTACACGACGCTCTTCCGATCTYCACAGACAACAAGAGGMCCGAG |
| TRAV12_2 both 12-4 | ACACTCTTTCCCTACACGACGCTCTTCCGATCTGTTATGCAATGGAAATTCCATAACCC |
| TRAV13 | ACACTCTTTCCCTACACGACGCTCTTCCGATCTCCTTGGTTCTGCAGGAGGSGGA |
| TRAV14_1 TRAV14-1 and TRAV14-2 | ACACTCTTTCCCTACACGACGCTCTTCCGATCTCAGCACTTTTRACTACTTCCCATG |
| TRAV14_3 TRAV14-3 | ACACTCTTTCCCTACACGACGCTCTTCCGATCTGAGAACAGTGCTTTTGACTACTTCC |
| TRAV15_not3 | ACACTCTTTCCCTACACGACGCTCTTCCGATCTTAGTGGAGAGATGGTTTTSCTTATTY |
| TRAV15_3 | ACACTCTTTCCCTACACGACGCTCTTCCGATCTGGACAGAAATTAACTCAGGTCCAAC |
| TRAV16 | ACACTCTTTCCCTACACGACGCTCTTCCGATCTGACAATGGACTGTGTGTATGAAACC |
| TRAV17 | ACACTCTTTCCCTACACGACGCTCTTCCGATCTTCCGTGGACCAGCCTGATGCTC |
| TRAV18 | ACACTCTTTCCCTACACGACGCTCTTCCGATCTAGAGTCCTCGGTTTCTGAGTATCC |
| TRAV19 | ACACTCTTTCCCTACACGACGCTCTTCCGATCTACTATTTTGCCTGGTACAAAAAATACC |
| TRAV20 | ACACTCTTTCCCTACACGACGCTCTTCCGATCTCAACTGCAGTTATACGAATGCAGC |
| TRAV21 | ACACTCTTTCCCTACACGACGCTCTTCCGATCTGGAAACGAGTACATCTATTGGTACC |
| TRAV22 | ACACTCTTTCCCTACACGACGCTCTTCCGATCTGATCTGTAGTTACCAAGACAGAGC |
| TRAV23 | ACACTCTTTCCCTACACGACGCTCTTCCGATCTATTCAAGATCCGCTACAGACAGCC |
| TRBV1 | ACACTCTTTCCCTACACGACGCTCTTCCGATCTGCATCTTGAAGAATTCCCAGTATCC |
| TRBV2 | ACACTCTTTCCCTACACGACGCTCTTCCGATCTCCACAATGCTATGTATTGGTATAGAC |
| TRBV3 | ACACTCTTTCCCTACACGACGCTCTTCCGATCTAGCAGATGGAGTTTCTGGTTAATTTC |
| TRBV4 | ACACTCTTTCCCTACACGACGCTCTTCCGATCTTAAGAAATTGCTGAAGATTATGTTTAGC |
| TRBV5 | ACACTCTTTCCCTACACGACGCTCTTCCGATCTAACATCTGGGACATAATGCTATGTAC |
| TRBV6 | ACACTCTTTCCCTACACGACGCTCTTCCGATCTTTTTTGTCACATTTTCATACCAACAGAC |
| TRBV7 | ACACTCTTTCCCTACACGACGCTCTTCCGATCTAATAGTCCCTTCTCTCTGGACACC |
| TRBV8 | ACACTCTTTCCCTACACGACGCTCTTCCGATCTCATGCTTTGGTACCGACAAGATCC |
| TRBV9 | ACACTCTTTCCCTACACGACGCTCTTCCGATCTACTGACAGAACTCAAGATGAGGCC |
| TRBV10 | ACACTCTTTCCCTACACGACGCTCTTCCGATCTCAGGGGTTGAAGCTGATCCATTAC |
| TRBV11 | ACACTCTTTCCCTACACGACGCTCTTCCGATCTCCCCTTTACTGGTATCGACAGATC |
| TRBV12 | ACACTCTTTCCCTACACGACGCTCTTCCGATCTAGTCCAACAGTTTGATGACTATCACT |
| TRBV13 | ACACTCTTTCCCTACACGACGCTCTTCCGATCTYATGTACTGGTATCGGCAGGACAC |
| TRBV14 | ACACTCTTTCCCTACACGACGCTCTTCCGATCTGTGACCCTATTTCTGGACATGATAC |
| TRBV15 | ACACTCTTTCCCTACACGACGCTCTTCCGATCTGAAGTGTGAGCCAGTTTCAGGCC |
| TRBV16 | ACACTCTTTCCCTACACGACGCTCTTCCGATCTGTGACAGGGAAGGGACAAGAAGC |
| TRBV17 | ACACTCTTTCCCTACACGACGCTCTTCCGATCTGTTCTCAGACTATGAATCATGATACC |
| TRBV18 | ACACTCTTTCCCTACACGACGCTCTTCCGATCTAACACAGTCATGTTATATGGTATAGTC |
| TRBV19 | ACACTCTTTCCCTACACGACGCTCTTCCGATCTGAAAACGATCTTCAAAAAGGCGATC |
| TRBV20 | ACACTCTTTCCCTACACGACGCTCTTCCGATCTGGAGTGTCAAGCTGTGGGTTTTC |
| TRBV21 | ACACTCTTTCCCTACACGACGCTCTTCCGATCTTACGTTTATTGGTACTACAAGAAACC |
| TRBV22 | ACACTCTTTCCCTACACGACGCTCTTCCGATCTAGCTCATGACAGAACAGACTTGTTC |
| TRBV23 | ACACTCTTTCCCTACACGACGCTCTTCCGATCTTAAAGTTCCTGATTTACTTTCAGAATC |
| TRBV24 | ACACTCTTTCCCTACACGACGCTCTTCCGATCTGTATCAACAGAACCAGAAACAAGAC |
| TRBV25 | ACACTCTTTCCCTACACGACGCTCTTCCGATCTTTTCTTTGTATTTACAGAATCAAGAAGC |
| TRBV26 | ACACTCTTTCCCTACACGACGCTCTTCCGATCTTGTATTCTGGTATCAACAAAATAAGAAC |
| TRBV27 | ACACTCTTTCCCTACACGACGCTCTTCCGATCTCAAAATGTATTGGTATCAGCAAGACC |
| TRBV28 | ACACTCTTTCCCTACACGACGCTCTTCCGATCTCATAGATGAGACTCATGTCTGTAATC |

| TRBV29 | ACACTCTTTCCCTACACGACGCTCTTCCGATCTAAACAATGTACTGGTATCGACAAGAC | | | | | | |
| TRBV30 | ACACTCTTTCCCTACACGACGCTCTTCCGATCTACAGAGCTTGATGCTCATGGCAAC | | | | | | |
| TRBV31 | ACACTCTTTCCCTACACGACGCTCTTCCGATCTTGCCGAGATCAAGGCTGTGGGC | | | | | | |

| Sequence Name | Sequence |
|---|---|
| macfas_TRA_1 | TCG TGG GCT CGG AGA TGT GTA TAA GAG ACA GAG AGT CTG TGA CCC AGC TTG ACA GCC A |
| macfas_TRA_2 | TCG TGG GCT CGG AGA TGT GTA TAA GAG ACA GAG AAA GAA CAG CTC CCT GCA CAT CAC A |
| macfas_TRA_3 | TCG TGG GCT CGG AGA TGT GTA TAA GAG ACA GAG AAA GGA TCC CAG CCT GAA GAC TCA G |
| macfas_TRA_4 | TCG TGG GCT CGG AGA TGT GTA TAA GAG ACA GAG AAA TAC TTT CAA GAA CTG CTT GGA A |
| macfas_TRA_5 | TCG TGG GCT CGG AGA TGT GTA TAA GAG ACA GAG AAA TCC TTC AGT CTC AAG ATC TCA G |
| macfas_TRA_6 | TCG TGG GCT CGG AGA TGT GTA TAA GAG ACA GAG AAC TGC ACT TAC AGC AAC AGT GCT T |
| macfas_TRA_7 | TCG TGG GCT CGG AGA TGT GTA TAA GAG ACA GAG CCA CAT ACC GTA AAG AAA CCA C |
| macfas_TRA_8 | TCG TGG GCT CGG AGA TGT GTA TAA GAG ACA GAG AGC TCC AGA TGA AAG ACT CTG CCT C |
| macfas_TRA_9 | TCG TGG GCT CGG AGA TGT GTA TAA GAG ACA GAG GAT GCT GAG TAC TTC TGT GCT G |
| macfas_TRA_10 | TCG TGG GCT CGG AGA TGT GTA TAA GAG ACA GAG CAG CAG CTG GAG CAG AGT CCT CGG T |
| macfas_TRA_11 | TCG TGG GCT CGG AGA TGT GTA TAA GAG ACA GAG TCA GCA ACG TAT TTC TGT GCA A |
| macfas_TRA_12 | TCG TGG GCT CGG AGA TGT GTA TAA GAG ACA GAG TGG ACT CAG CAG TAT ACT TCT GTG C |
| macfas_TRA_13 | TCG TGG GCT CGG AGA TGT GTA TAA GAG ACA GAG GTT CAA ATA TGG CTA AGA GGC AAG G |
| macfas_TRA_14 | TCG TGG GCT CGG AGA TGT GTA TAA GAG ACA GAG CAT ATG GAG CAG GGA GAC AAA CAT T |
| macfas_TRA_15 | TCG TGG GCT CGG AGA TGT GTA TAA GAG ACA GAG AAT CTC TTC TGG TAT GTC CAG TAC C |
| macfas_TRA_16 | TCG TGG GCT CGG AGA TGT GTA TAA GAG ACA GAG CTG TAA GCC AGC ATA ACC ACC ATG T |
| macfas_TRA_17 | TCG TGG GCT CGG AGA TGT GTA TAA GAG ACA GAG TGT AAC TCT CAA CTG CAG TTA TGA A |
| macfas_TRA_18 | TCG TGG GCT CGG AGA TGT GTA TAA GAG ACA GAG AAG GCG CCC ATC TTC CTG ATG ATA T |
| macfas_TRA_19 | TCG TGG GCT CGG AGA TGT GTA TAA GAG ACA GAG AAA GCG ACA GCG TCA CAC TGA ACT G |
| macfas_TRA_20 | TCG TGG GCT CGG AGA TGT GTA TAA GAG ACA GAG AGA ATG GCC TCT CTG ACA ATC ACT G |
| macfas_TRA_21 | TCG TGG GCT CGG AGA TGT GTA TAA GAG ACA GAG CAA ATG GAG CAG TGA AGC AGG AGG G |
| macfas_TRA_22 | TCG TGG GCT CGG AGA TGT GTA TAA GAG ACA GAG AGA CAT TAT ATG GCT TAC ACT GGT A |
| macfas_TRA_23 | TCG TGG GCT CGG AGA TGT GTA TAA GAG ACA GAG ATC AGG GTT ATT CTA AGT CAA ATG C |
| macfas_TRA_24 | TCG TGG GCT CGG AGA TGT GTA TAA GAG ACA GAG GGA GAC AAC TTG GTT CTC AAC TGC A |
| macfas_TRA_25 | TCG TGG GCT CGG AGA TGT GTA TAA GAG ACA GAG CCT CTC TGA TCA TCA CAG AAG ACA G |
| macfas_TRA_26 | TCG TGG GCT CGG AGA TGT GTA TAA GAG ACA GAG AGT CCA GAA AGG AGG GAT TTC AAT T |
| macfas_TRA_27 | TCG TGG GCT CGG AGA TGT GTA TAA GAG ACA GAG ACT TCT TCA AGC ACA TTT AAC ACC T |
| macfas_TRA_28 | TCG TGG GCT CGG AGA TGT GTA TAA GAG ACA GAG CTG GCT GCA ACA GCA TCC AGG AGG A |
| macfas_TRA_29 | TCG TGG GCT CGG AGA TGT GTA TAA GAG ACA GAG AAT GCT AAA CAT GTC TCC CTG CAT A |
| macfas_TRA_30 | TCG TGG GCT CGG AGA TGT GTA TAA GAG ACA GAG ACA TAT ATT CTT TCA AAT ACG GAC C |
| macfas_TRA_31 | TCG TGG GCT CGG AGA TGT GTA TAA GAG ACA GAG CCA TCT GCC CTT GTG AGC GAC T |
| macfas_TRA_32 | TCG TGG GCT CGG AGA TGT GTA TAA GAG ACA GAG TCA ACA AGG AGA AGA GGA TCC TCA G |
| macfas_TRA_33 | TCG TGG GCT CGG AGA TGT GTA TAA GAG ACA GAG ACA AGA CAG CCA AAC ATT TCT CTC T |
| macfas_TRA_34 | TCG TGG GCT CGG AGA TGT GTA TAA GAG ACA GAG ATC ATG TGG TAC CAA CAG TTT CCC A |
| macfas_TRA_35 | TCG TGG GCT CGG AGA TGT GTA TAA GAG ACA GAG CCC TGA GAG GGC AGC TCT AAC ATT A |
| macfas_TRA_36 | TCG TGG GCT CGG AGA TGT GTA TAA GAG ACA GAG AGC ACA GCT CGA TAG AGC CAG CCA G |
| macfas_TRA_37 | TCG TGG GCT CGG AGA TGT GTA TAA GAG ACA GAG ACT CTT AAA CAG AGT TTG TTT CAT A |
| macfas_TRA_38 | TCG TGG GCT CGG AGA TGT GTA TAA GAG ACA GAG TCT CAG CAG GGA CGA TAC AAC ATG A |
| macfas_TRA_39 | TCG TGG GCT CGG AGA TGT GTA TAA GAG ACA GAG CTG CTG AAG GTC CTA CAT TCC TGA T |
| macfas_TRA_40 | TCG TGG GCT CGG AGA TGT GTA TAA GAG ACA GAG TCA TGA CTT TCA GTG AGA ACA CAA A |
| macfas_TRA_41 | TCG TGG GCT CGG AGA TGT GTA TAA GAG ACA GAG CAC ATT TCC TCC TCC CAG ACC ACA G |
| macfas_TRA_42 | TCG TGG GCT CGG AGA TGT GTA TAA GAG ACA GAG CAG GTA TCA GAC TCA GCC GTG TAC T |
| macfas_TRA_43 | TCG TGG GCT CGG AGA TGT GTA TAA GAG ACA GAG TGA CGC AGA GTC CTG AGG CCC T |
| macfas_TRA_44 | TCG TGG GCT CGG AGA TGT GTA TAA GAG ACA GAG AGA GTC AAT ATT CAG TGA GCT TCC A |
| macfas_TRAV11_3 | TCG TGG GCT CGG AGA TGT GTA TAA GAG ACA GAGGTTTGAATATCCCGACCTCTCATC |
| macfas_TRAV14-1 | TCG TGG GCT CGG AGA TGT GTA TAA GAG ACA GAGTCTGGTACAAGTAGCCCAGGAGTGG |
| macfas_TRAV46 | TCG TGG GCT CGG AGA TGT GTA TAA GAG ACA GAG TCC TTA AGC ATG TGC AGC CAA GAG A |
| macfas_TRAV8-6 | TCG TGG GCT CGG AGA TGT GTA TAA GAG ACA GAGGAACGACACGCTGAGTACTTCTGTG |
| macfas_TRB_1 | TCG TGG GCT CGG AGA TGT GTA TAA GAG ACA GAG AAA CAA ACA GAA AGC AAA GAT GGA T |
| macfas_TRB_2 | TCG TGG GCT CGG AGA TGT GTA TAA GAG ACA GAG AAA AGA TGG ATT GTA CCC CCG A |
| macfas_TRB_3 | TCG TGG GCT CGG AGA TGT GTA TAA GAG ACA GAG AAA CAA GAG GAC AGC AAG TGA CAC T |
| macfas_TRB_4 | TCG TGG GCT CGG AGA TGT GTA TAA GAG ACA GAG AAA CAG CCA CTC TGC AAT GCT ATC C |
| macfas_TRB_5 | TCG TGG GCT CGG AGA TGT GTA TAA GAG ACA GAG AAA CAT GCA CTG AGT CAA CAC AGT C |
| macfas_TRB_6 | TCG TGG GCT CGG AGA TGT GTA TAA GAG ACA GAG AAA CCC CAA AGC ACC TGA TCA CAG C |
| macfas_TRB_7 | TCG TGG GCT CGG AGA TGT GTA TAA GAG ACA GAG CGG ATA TTC AGA GAG GAG ACC T |
| macfas_TRB_8 | TCG TGG GCT CGG AGA TGT GTA TAA GAG ACA GAG CGT GAG CAT CTG GGA CAT GAT T |
| macfas_TRB_9 | TCG TGG GCT CGG AGA TGT GTA TAA GAG ACA GAG AAA GAG GAG AAA GGA TGT AGC TCT C |
| macfas_TRB_10 | TCG TGG GCT CGG AGA TGT GTA TAA GAG ACA GAG AAA GAT CAC TCT GGA ATG TTC T |
| macfas_TRB_11 | TCG TGG GCT CGG AGA TGT GTA TAA GAG ACA GAG AAA GAA TGT TCT GGT ATC GAC AAG A |
| macfas_TRB_12 | TCG TGG GCT CGG AGA TGT GTA TAA GAG ACA GAG GAG TTC AAG TTC TTG ATT TCC T |
| macfas_TRB_13 | TCG TGG GCT CGG AGA TGT GTA TAA GAG ACA GAG AAA GAT TTT AAC AAT GAA GCA GAC A |
| macfas_TRB_14 | TCG TGG GCT CGG AGA TGT GTA TAA GAG ACA GAG AAA CCC AAG CAG GGA TGT CTG TCA A |
| macfas_TRB_15 | TCG TGG GCT CGG AGA TGT GTA TAA GAG ACA GAG AAA GGA AGG ACA GAA TGT GAC CCT G |
| macfas_TRB_16 | TCG TGG GCT CGG AGA TGT GTA TAA GAG ACA GAG AAA GGA ATT CAT TCT ACA GTG TTC C |
| macfas_TRB_17 | TCG TGG GCT CGG AGA TGT GTA TAA GAG ACA GAG AAA GGA CAC AGT CAT GTT TAT TGG T |
| macfas_TRB_18 | TCG TGG GCT CGG AGA TGT GTA TAA GAG ACA GAG AAA GGA GTT CTC AGG ATA TGT GCC A |
| macfas_TRB_19 | TCG TGG GCT CGG AGA TGT GTA TAA GAG ACA GAG AAA TCA GTC TCA GCC CCA TGC CTG G |
| macfas_TRB_20 | TCG TGG GCT CGG AGA TGT GTA TAA GAG ACA GAG TGA GAG GAC AGC AAG CGA CAT T |
| macfas_TRB_21 | TCG TGG GCT CGG AGA TGT GTA TAA GAG ACA GAG AAA TGG GAC AAG AAG TAA CTA TGA G |
| macfas_TRB_22 | TCG TGG GCT CGG AGA TGT GTA TAA GAG ACA GAG AAA TGG GAC AAG CAG TGA CTC TCA G |
| macfas_TRB_23 | TCG TGG GCT CGG AGA TGT GTA TAA GAG ACA GAG AAA TTC CAG GTC CTG AAG ACA GGA C |
| macfas_TRB_24 | TCG TGG GCT CGG AGA TGT GTA TAA GAG ACA GAG AAA TTT CTG ATA TAC TTT CTG AGA G |
| macfas_TRB_25 | TCG TGG GCT CGG AGA TGT GTA TAA GAG ACA GAG AAC AAC AAG TCT CCG ATA GAT GAT T |
| macfas_TRB_26 | TCG TGG GCT CGG AGA TGT GTA TAA GAG ACA GAG AAC ACC AAT GGA GTA GAA GAG CAG C |
| macfas_TRB_27 | TCG TGG GCT CGG AGA TGT GTA TAA GAG ACA GAG AAC CTC TGT GAA GAT CGA GTG CCG T |
| macfas_TRB_28 | TCG TGG GCT CGG AGA TGT GTA TAA GAG ACA GAG AAC CTG AAG TCA CCC AGA CTC CCA G |
| macfas_TRB_29 | TCG TGG GCT CGG AGA TGT GTA TAA GAG ACA GAG AAG AAA TTG CTG AAG ATA ATG TTT A |
| macfas_TRB_30 | TCG TGG GCT CGG AGA TGT GTA TAA GAG ACA GAG AAG TCT TTG AAA TGT GAA CAA C |
| macfas_TRB_31 | TCG TGG GCT CGG AGA TGT GTA TAA GAG ACA GAG AAT CAC CCA GAG CCC AAG ACA CAA G |
| macfas_TRB_32 | TCG TGG GCT CGG AGA TGT GTA TAA GAG ACA GAG ACC TTG ACG TGT CAC CAG ACT TGG A |
| macfas_TRB_33 | TCG TGG GCT CGG AGA TGT GTA TAA GAG ACA GAG AGA TGA TTC ACA GTT GCC TAA GGA T |
| macfas_TRB_34 | TCG TGG GCT CGG AGA TGT GTA TAA GAG ACA GAG AGA TGC TCT CCT ATC TCT GGG CAC A |
| macfas_TRB_35 | TCG TGG GCT CGG AGA TGT GTA TAA GAG ACA GAG AGG GCC CAG AGT TTC TGA CTT ACT T |
| macfas_TRB_36 | TCG TGG GCT CGG AGA TGT GTA TAA GAG ACA GAG AGG TGT GAT CCA ATT TCG GGT CAT G |
| macfas_TRB_37 | TCG TGG GCT CGG AGA TGT GTA TAA GAG ACA GAG ATG TAC TGG TAT CGA CAA GAC CCA G |
| macfas_TRB_38 | TCG TGG GCT CGG AGA TGT GTA TAA GAG ACA GAG ATG TCC TGG TAT CGA CAA GAC CCA G |
| macfas_TRB_39 | TCG TGG GCT CGG AGA TGT GTA TAA GAG ACA GAG CAG GTG TGA TCC AAT TTC AGG TCA T |
| macfas_TRB_40 | TCG TGG GCT CGG AGA TGT GTA TAA GAG ACA GAG CCA GAC AAA TCA GGG CTG CCC AGT G |
| macfas_TRB_41 | TCG TGG GCT CGG AGA TGT GTA TAA GAG ACA GAG CTG GTA TCG ACA AGA CCC AGG CAT G |
| macfas_TRB_42 | TCG TGG GCT CGG AGA TGT GTA TAA GAG ACA GAG TTA GGA ACC ATT ATT CAA TGT TCT G |
| macfas_TRBV1_3 | TCG TGG GCT CGG AGA TGT GTA TAA GAG ACA GAG CTC TCG CTT ATG CCT TCA TGT GGT C |
| macfas_TRBV17_1 | TCG TGG GCT CGG AGA TGT GTA TAA GAG ACA GAG GCC ATA TAT CTC CGC AGT AGC GAT G |
| macfas_TRBV_6_5 | TCG TGG GCT CGG AGA TGT GTA TAA GAG ACA GAG TCC CCA ATG GCT ACA ATG TCT TTA C |